\documentstyle[12pt,times,twoside,psfig]{article}
\newcommand{\beq}{\begin{equation}}
\newcommand{\eeq}{\end{equation}}
\newcommand{\bea}{\begin{eqnarray}}
\newcommand{\eea}{\end{eqnarray}}

\begin{document}

\begin{titlepage}

{\noindent
{\Large \bf
LA-UR-98-0418, LANL, Los Alamos (1998);\\
submitted to {\em Nuclear Science and Engineering}\\
}}

\vspace*{1cm}

\begin{center}
{\LARGE
Cascade-Exciton Model Analysis of Nucleon-Induced \\
Fission Cross Sections of Lead and Bismuth \\
at Energies from 45 to 500 MeV\\}
\vspace*{1.2cm}
{A.~V.~Prokofiev} \\
{\it V. G. Khlopin Radium Institute,
2nd Murinsky av. 28, St. Petersburg 194021, Russia} \\
\vspace*{.3cm}
{and}\\
\vspace*{.3cm}
{S.~G.~Mashnik and A.~J.~Sierk}\\
{\it T-2, Theoretical Division, Los Alamos National Laboratory,
Los Alamos, NM 87545}\\
\vspace{2.0cm}
\end{center}
\vspace*{10pt}
\noindent{
\hspace*{-2pt}
{\bf Abstract}---{\it
An extended version of the Cascade-Exciton Model (CEM) of nuclear
reactions is applied to analyze nucleon-induced fission cross
sections for $^{209}$Bi and $^{208}$Pb nuclei in the
45--500 MeV energy range.
The available data on linear momentum transfer are analyzed
as well. The results are compared with analytical approximations resulting
from a comparative critical analysis of all available experimental data.
Systematic discrepancies between calculations and experimental data are
revealed. A modification of the CEM is proposed, which
significantly improves the model predictions for projectile energies
above 100 MeV.}
}

\end{titlepage}

\newpage

\begin{center}
{\large I. INTRODUCTION} \\
\end{center}

A large amount of nuclear reaction data,
including fission cross sections at intermediate energies is
required for applications, e.g., for
accelerator transmutation of waste (ATW) for elimination of
long-lived radioactive wastes with a spallation source,
accelerator-based conversion (ABC) aimed to complete the destruction of weapon
plutonium,
accel\-erator-driven energy production (ADEP) which proposes to derive
fission energy from thorium with concurrent destruction of the long-lived waste
and without the production of weapon-usable material, for accelerator
production of tritium (APT) etc. (see the Proceedings of the
international conferences on Accelerator-Driven Transmutation Technology
held in Kalmar~\cite{kalmar96} and
Dubna~\cite{dubna96} and references therein).
Experiments to measure these data are costly and there are a limited
number of facilities available to make such measurements.
Therefore reliable models are required to provide the necessary data.

During the last two decades, several versions of the Cascade-Exciton
Model (CEM)~\cite{cem} of nuclear reactions have been developed
at JINR, Dubna (for an overview, see ~\cite{cemvar94}).
A large variety of experimental data
for reactions induced by nucleons~\cite{cemnuclons}, pions~\cite{cempions},
and photons~\cite{cemphoto} has been analyzed in the framework of the CEM
and the general validity of this approach has been confirmed. The recent
{\it International Code Comparison for Intermediate Energy Nuclear
Data}~\cite{nea94a} has shown that the CEM adequately describes nuclear
reactions at intermediate energies and has one of the best predictive
powers for double differential cross sections of secondary nucleons as
compared to other available models (see Tabs. 5 and 6 in the
Report~\cite{nea94a} and Fig.~7 in Ref.~\cite{cemphys}).
In the last few years, the CEM has been extended~\cite{pit95}  to calculate
hadron-induced spallation and used to study~\cite{report97}--\cite{konshin}
about 700 reactions induced by protons from 10 MeV to 5 GeV
on nuclei from Carbon to Uranium.

A detailed description of the CEM may be found in Ref.~\cite{cem}
and of its extended version, as realized in the code CEM95, in
Refs.~\cite{pit95,report97};
therefore, we mention here only its basic assumptions. The CEM
assumes that reactions occur in three stages. The first stage is
the intranuclear cascade in which primary and
secondary particles can be rescattered
several times prior to absorption by, or escape from the nucleus.
The cascade stage of the interaction is described by the standard
version of the Dubna intranuclear cascade model (ICM)~\cite{book}.
The excited residual nucleus remaining after the emission of the
cascade particles determines the particle-hole configuration that is
the starting point for the second, preequilibrium stage of the
reaction. The subsequent relaxation of the nuclear excitation is
treated by an extension of the Modified Exciton model
(MEM)~\cite{mem} of preequilibrium decay which also
includes the description of the equilibrium evaporative third stage of
the reaction.

Recently, the CEM has been extended by taking into account the competition
between particle emission and fission at the compound nucleus
stage~\cite{acta2} and a more realistic calculation of nuclear level
density~\cite{acta1}. An earlier extended version of the CEM, as realized
in the code CEM92, was used by Konshin~\cite{konshin}
to calculate
nucleon-induced fission cross sections for actinides in the
energy region from 100 MeV to 1 GeV.
Nevertheless, previously the CEM has not been applied to study
fission cross sections for pre-actinides; therefore, its
predictive power and applicability to evaluate arbitrary
fission cross sections was unknown.

In order to investigate the applicability of the CEM to fission cross
sections and hoping to learn more about
intermediate-energy fission and to identify possible improvements to
the CEM and other models to improve their predictive power,
we use here an extended version of the CEM
(see the recent report~\cite{report97} and references therein),
as realized in the code CEM95~\cite{cem95manual},
to perform a detailed analysis of proton- and  neutron-induced fission cross
sections of $^{209}$Bi and $^{208}$Pb nuclei and of
the linear momentum transfer to the fissioning nuclei
in the 45--500 MeV energy range.

The $^{208}$Pb and $^{209}$Bi target nuclei
and this energy interval were chosen for this study for the
following reasons:

\begin{itemize}
\item
new experimental information on $^{208}$Pb(n,f) and $^{209}$Bi(n,f)
reactions has been recently obtained at these incident energies
(see~\cite{staples}-\cite{eismont96b} and references therein),
\item
the reactions under study have practical importance  in connection
with concepts of accelerator-driven transmutation and tritium
production technologies, which may include the use
as a neutron source of massive
lead or lead-bismuth targets irradiated by proton beams,
\item
cross sections for these reactions are widely used as standards in the
intermediate energy region (e.g., \cite{conde94,smirnov97}), with cross
sections for other targets being normalized to them,
\item
fissioning nuclei in this mass region are close to the doubly magic nucleus
$^{208}$Pb, where one would expect the clearest manifestation of shell
effects, which are important for the development of any model,
\item
fission barriers of nuclei in this mass region do not have the
double-humped structure which appears in the actinide nuclei and makes
calculations for them more complicated and uncertain.
\end{itemize}

\begin{center}
{\large II. CALCULATIONAL TECHNIQUE} \\
\end{center}

As the fission cross sections for nuclei under study constitute a small
part of the total reaction cross section, the statistical weight
method~\cite{acta2,barashenkov74}
is used to calculate fission cross sections instead of the
straightforward Monte-Carlo technique. It allows us to obtain statistical
uncertainties of the calculated fission cross sections, as a rule,
not more than 5--10\% with only 3000 inelastic events in each calculation.

In the course of the calculations, the characteristics of the fissioning nuclei
(the charge, the mass, the excitation energy, all components of the linear and
angular momenta, and the statistical weight of the event) are saved, and later
used for the construction of distributions of fission
events with respect to specified parameters as well as for the calculation of
the
corresponding average values. Examples of these distributions are
shown in Figs.~1 and 2.

Fig.~1 represents the distributions of calculated longitudinal linear momentum
transfer (LMT) to the fissioning nuclei in  the $^{209}$Bi(p,f) reaction
for incident proton energies of 45 MeV (a), 73 MeV (b), 96 MeV (c),
and 160 MeV (d). Comparing
these distributions allows one to follow the changes in the reaction mechanism
with increasing incident particle energy. The calculation predicts that
most nuclei undergoing fission induced by 45 MeV protons possess a linear
momentum equal to that of the incident proton (Fig.~1a), i.e., the full
LMT mode predominates, which corresponds to
fission from the compound nucleus. (Hereinafter the term ``compound nucleus"
implies the nucleus with the charge and the mass equal to the sum of the
corresponding characteristics of the target nucleus and of
the incident particle.)
With increasing incident proton energy, the share of the events with the full
LMT falls quickly (Fig.~1b). For proton energies of 100 MeV or more (Fig.~1c,
d)
the fission from the compound nucleus practically
``dies out"; the  maximum in the distribution becomes displaced to the region
of
incomplete LMT and the distribution width increases. A comparison of the LMT
calculations with experimental data is discussed in Sect.~III.

Fig.~2 shows the distributions of the charge of the fissioning nuclei (a, b),
mass (c, d), and excitation energy (e, f) in the $^{209}$Bi(p,f) reaction for
proton energies of 45 and 160 MeV. The curves reflect the same tendency as
discussed above. At 45 MeV, fission from the compound nucleus ($^{210}$Po)
predominates. On the contrary, at 160 MeV the distributions become displaced
and
their shape approaches symmetry.

In Fig.~3, we show the dependence of the average charge (a), mass (b) and
excitation energy (c) of the fissioning nuclei in the $^{209}$Bi(p,f) and
$^{209}$Bi(n,f) reactions as functions of
the incident particle energy in the 45--160 MeV range.
The energy dependence of the average mass and
excitation energy of the fissioning nuclei are nearly the same for these two
reactions. On the other hand, the average charge of the fissioning nuclei for
the (p,f) reaction decreases faster than for the
(n,f) one. This  can be qualitatively explained by noticing that there is a
significant probability that the incident nucleon leaves the nucleus
after an inelastic scattering during the intranuclear cascade stage of
the reaction.

The results presented in this and subsequent sections are obtained with
input parameters of CEM95 as discussed in Sect.~V.\\

\newpage

\begin{figure}[h!]
\centerline{
\psfig{figure=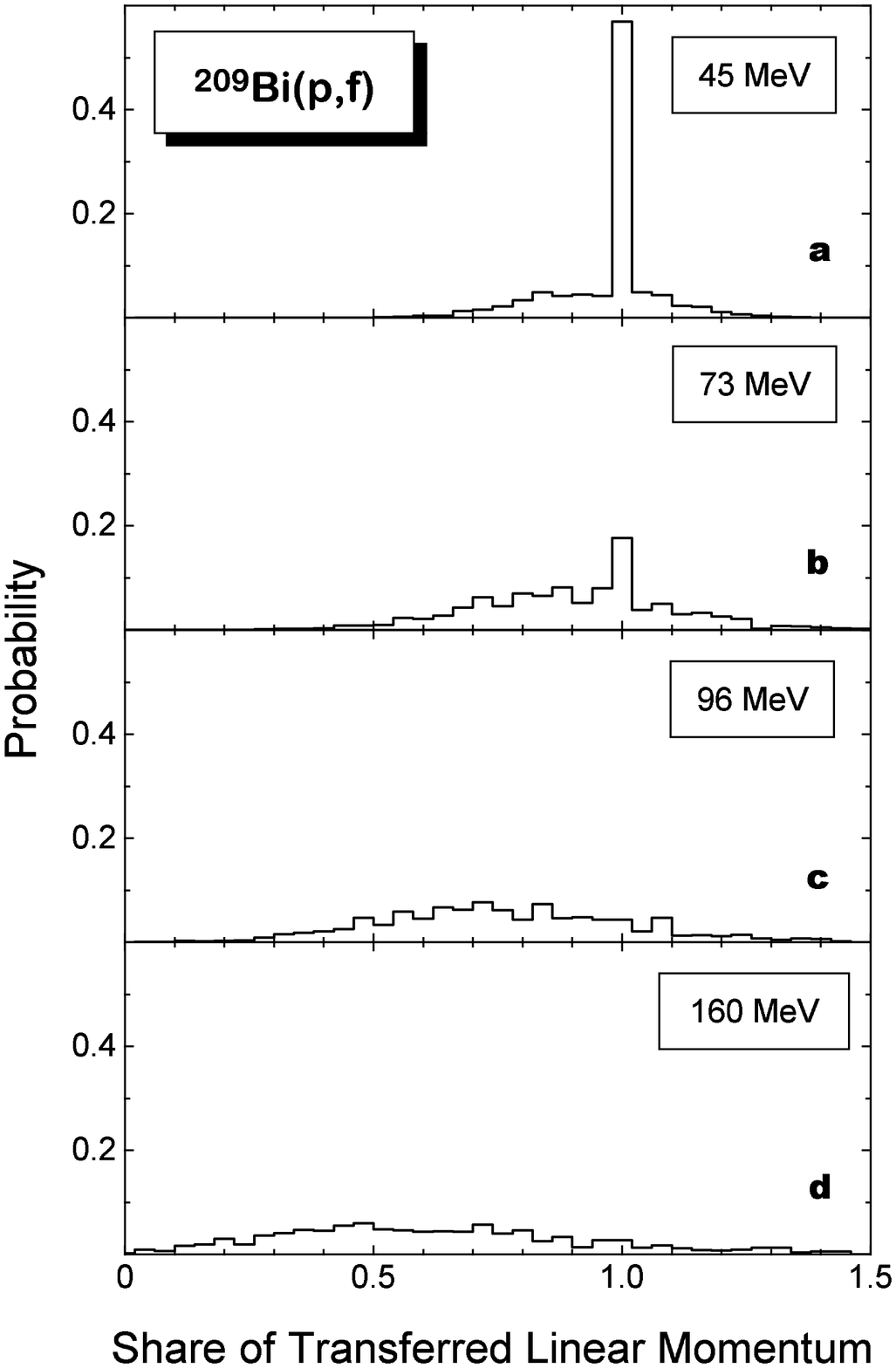,width=130mm,angle=0}}
\end{figure}

{\small
Fig.~1. Calculated distributions of longitudinal linear momentum transferred to
the
fissioning nuclei in the $^{209}$Bi(p,f) reaction for
incident proton energies of
45~MeV~(a), 73~MeV~(b), 96~MeV~(c), and 160~MeV~(d).
}

\newpage

\begin{figure}[h!]
\centerline{
\psfig{figure=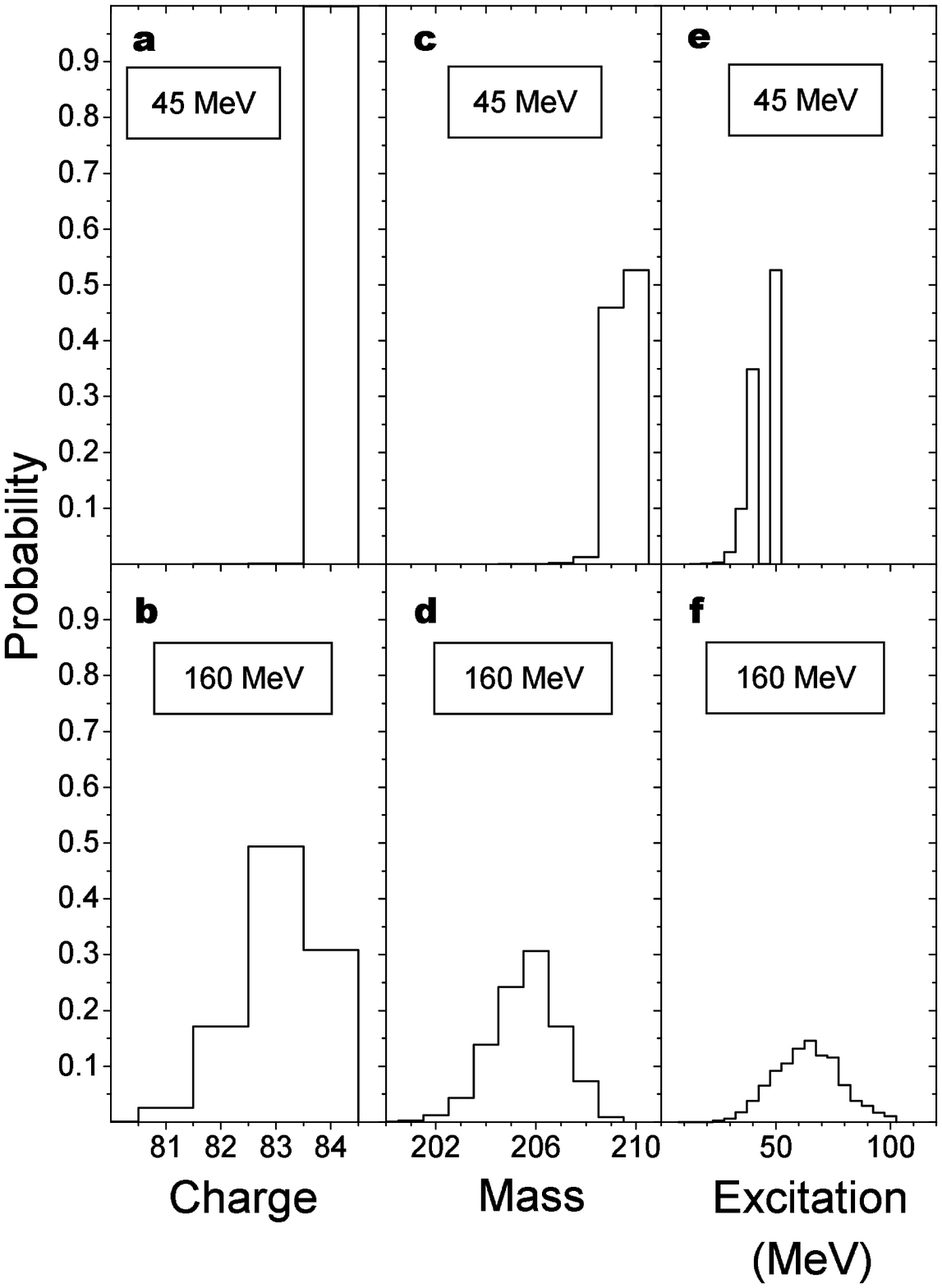,width=140mm,angle=0}}
\end{figure}

{\small
Fig.~2. Calculated distributions of
fissioning nuclei in the $^{209}$Bi(p,f) reaction vs.
charge~(a,b), mass~(c,d), and excitation energy~(e,f) for
incident proton energies of
45~MeV~(a,c,e) and 160~MeV~(b,d,f).
}

\newpage

\begin{figure}[h!]
\centerline{
\psfig{figure=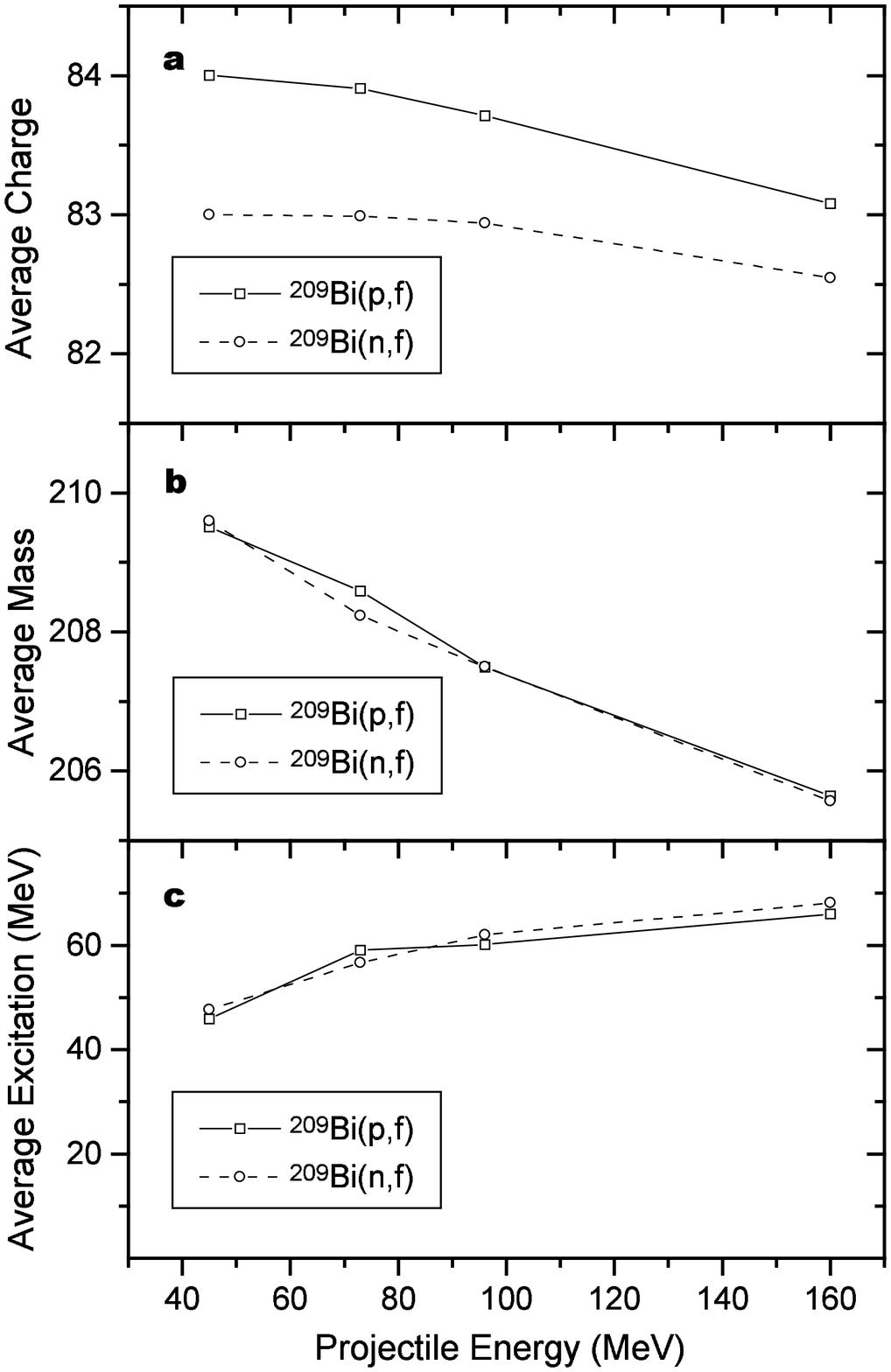,width=130mm,angle=0}}
\end{figure}

{\small
Fig.~3. Calculated average charge~(a), mass~(b), and excitation energy~(c)
of fissioning nuclei in the $^{209}$Bi(p,f) reaction
as a function of projectile energy.
}

\newpage
\begin{center}
{\large III. LINEAR MOMENTUM TRANSFER} \\
\end{center}

It has been previously shown that the momentum transferred in
nucleon reactions with actinide nuclei is better described by the CEM than
by other models (see details in Ref.~\cite{sasha93}).

Fig.~4 shows the calculated average LMT for the $^{209}$Bi(p,f)~(a),
$^{209}$Bi(n,f)~(b), $^{208}$Pb(p,f)~(c), and $^{208}$Pb(n,f)~(d)
reactions in
comparison with the sparse available experimental
data~\cite{kowalski63}--\cite{eismont97}.
A large spread exists between the $^{209}$Bi(p,f) LMT
data from different authors (Kowalski~\cite{kowalski63},
Stephan~\cite{stephan65}, Shigaev~\cite{shigaev78})
and even within the same experimental dataset~\cite{shigaev78}.
This may indicate the presence of some omitted systematic
uncertainties in that work~\cite{shigaev78}, which was performed with
solid-state nuclear track detectors. Possible sources of systematic
uncertainty may include a decrease of detection efficiency at back
angles, fluctuations of the position of the beam at the target, as well as
inhomogeneity of the target. We regard as more reliable the data from the
works~\cite{kowalski63,stephan65}
where fission fragments were registered by semiconductor detectors
in coincidence mode.  Judging from these data, one can
conclude that our present CEM95 calculation predicts correctly a decrease
with energy of the average LMT for the $^{209}$Bi(p,f) reaction but
underestimates its average value by 1--1.5 times the experimental uncertainty
for the highest energy.

To the best of our knowledge,
only one experimental work is available on LMT for the
(n,f) type of reactions~\cite{eismont97} (see Fig.~4b). In this work
the average LMT was obtained for the  $^{209}$Bi(n,f) reaction at 75
MeV by measuring the difference  between the kinetic energies of the fission
fragments registered by an ionization chamber in the forward and backward
hemispheres.
In contrast to the $^{209}$Bi(p,f) results, the
average LMT calculated  with the CEM95 code lies above the sole
experimental point by about 1.5 times the experimental uncertainty.

The works of Kowalski~\cite{kowalski63} and Stephan~\cite{stephan65}
also include measurements of LMT
distributions for the fissioning nuclei in the $^{209}$Bi(p,f) reaction at 156
and 96 MeV, respectively. To a first approximation, the distributions are
Gaussian-shaped, similar to the CEM95 predictions (see Fig.~1c, d). For
a quantitative comparison between the calculations and
experiment, we calculate the second moment (dispersion) of the LMT
experimental and calculated
distributions. In order to compare the  experimental dispersion
with the calculated one, it is necessary to introduce a correction to the
latter,
which takes into account the prompt neutron emission from the fission
fragments. We estimate this correction using the average
kinetic energy of the prompt fission neutrons
from Ref.~\cite{halpern} and the average number of
neutrons per fission from Ref.~\cite{gangrsky}.
The comparison between calculated and experimental results is shown in Table I.

Table I shows that there is good agreement between the experimental
data and our results corrected for the prompt neutron emission from the
fission fragments.

On the whole,
we conclude that the description of linear momentum transfer processes
provided by CEM95 does not contradict the available experimental data.

In Sect.~VI we discuss a sensitivity study of the calculated fission
cross sections to different input parameters of CEM95. We also study
 the sensitivity of the LMT description.
It is found that neither the average nor the dispersion of the LMT
varies beyond the statistical uncertainties of the
calculations for any reasonable variation of the input parameters.

\newpage

\begin{figure}[h!]
\centerline{
\psfig{figure=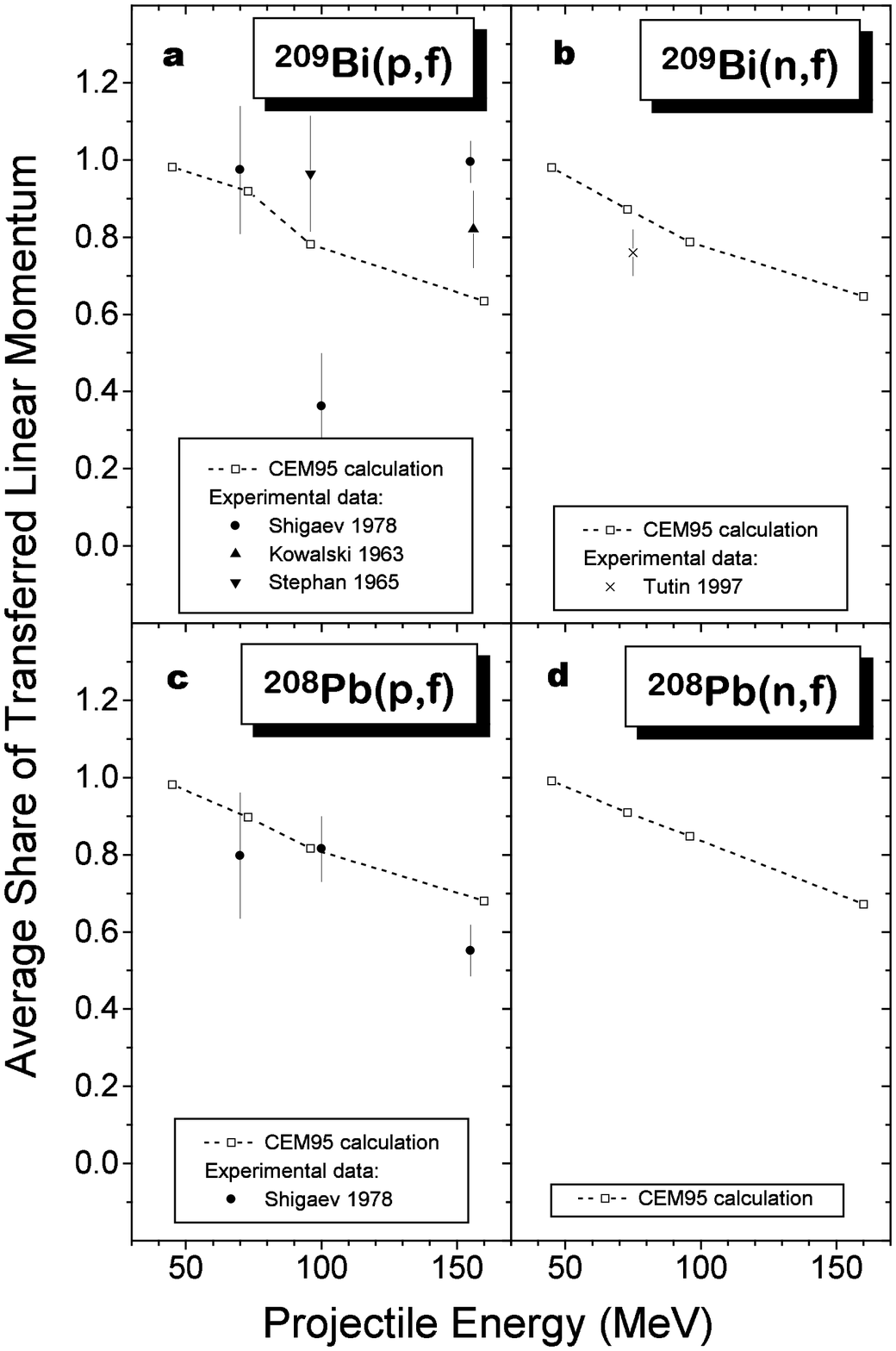,width=130mm,angle=0}}
\end{figure}

{\small
Fig.~4.
The average share of the linear momentum transferred to the
fissioning nuclei as a function of the projectile energy, for the following
reactions:
$^{209}$Bi(p,f)~(a), $^{209}$Bi(n,f)~(b), $^{208}$Pb(p,f)~(c),
and $^{208}$Pb(n,f)~(d).
The points are experimental data from
Refs.~\cite{kowalski63}--\cite{eismont97}. The dashed lines
and the open squares represent the present calculations.
}

\newpage
{\noindent This
insensitivity can be interpreted as follows. The momentum losses occur mainly
at the cascade stage of the reaction due to forward-peaked emission of
fast cascade particles.} As the excited nucleus
nears statistical equilibrium, the secondary particles become less
energetic and their emission becomes more isotropic in the
center-of-mass system. Thus the secondaries being emitted at the later reaction
stages take away less and less momentum. Since the free parameters
available in CEM95 relate only to these later reaction stages, it is natural
that
their variations have only a weak influence on the resulting LMT distributions.

\begin{center}
TABLE I\\
Dispersion of the LMT distribution in the  $^{209}$Bi(p,f) reaction \\
(in units of the incident proton linear momentum)
\end{center}

\begin{center}
\begin{tabular}{|c|c|c|c|c|}
\hline
Proton energy & CEM95 results & Correction due to & CEM95 &
Experimental data \\
(MeV) & (uncorrected) & prompt neutron & corrected & \\
 & & emission by & results &\\
 & & fission fragments & &\\
\hline
96 & 0.25 & 0.28 & 0.38 & $0.36\pm0.02$~\cite{stephan65}\\
160 & 0.32 & 0.21 & 0.38 & $0.38\pm0.02^{*}$~\cite{kowalski63}\\
\hline
\end{tabular}
\end{center}

$^{*}$ The experimental value was obtained at the proton energy of 156 MeV.\\

Furthermore, no notable differences are seen between the calculated LMT data
for incident protons and neutrons of the same energy.
The calculated results for
$^{208}$Pb and $^{209}$Bi target nuclei are also similar. Thus, CEM95 predicts
a
weak sensitivity of the LMT processes to variations of the incident particle
charge as well as to the target nucleus charge and mass.\\

\begin{center}
{\large IV. APPROXIMATING EXPERIMENTAL FISSION CROSS SECTIONS} \\
\end{center}

In contrast to the poor experimental database for LMT, the measurements of
fission cross sections for $^{208}$Pb and $^{209}$Bi are numerous although
their reliability varys greatly. Thus the most effective way to estimate the
quality of the model predictions is a comparison of the calculated results with
approximations built on the basis of the experimental data taking into
account differences in their systematic and statistical uncertainties.

Prokofiev and co-authors~\cite{sasha97}
performed a literature search and a comparative critical
analysis, ranking the experimental data on the (p,f) cross
sections for a wide range of nuclei from Ta to U in the energy region up to 30
GeV. We consider only the data for $^{208}$Pb and $^{209}$Bi up to 500 MeV.
An approximation for
the $^{209}$Bi(n,f) cross section was suggested by Smirnov and published
in Ref.~\cite{smirnov97}.
The approximation given below for the $^{208}$Pb(n,f) cross section
is obtained here and has not been published previously.

A list of publications containing the experimental data on the nucleon-induced
fission cross sections for $^{208}$Pb and $^{209}$Bi is given in Table II.

\begin{center}
TABLE II\\
References to publications containing experimental data on the
nucleon-induced fission cross sections for $^{208}$Pb and $^{209}$Bi at
energies up to 500 MeV
\end{center}

\begin{center}
\begin{tabular}{|c|c|c|}
\hline
Target nucleus & \multicolumn{2}{|c|} {Incident particle} \\
\hline
 & p & n \\
\hline
$^{209}$Bi & \cite{kowalski63,stephan65}, \cite{shigaev73}--\cite{flerov} &
\cite{staples}--\cite{eismont96b}, \cite{vorotnikov84,gold55} \\
$^{208}$Pb & \cite{shigaev73,okolovich,khodai,kon66,flerov} &
\cite{reut,gold55}, \cite{staples}--\cite{eismont96a} \\
\hline
\end{tabular}
\end{center}

\vspace*{0.5cm}
The experimental data from the publications listed in Table II were subjected
to
a critical comparative analysis
\cite{sasha97} in order to establish how reliably the quantities defining
fission cross sections (incident particle flux, target thickness, fission
fragment detection efficiency, etc.)~were measured. As a result of this
analysis, ``reliability grades" were assigned to every dataset based
on the following criteria:

\begin{itemize}
\item
the 1st (highest) grade was assigned to a dataset if the analysis
did not reveal any omitted or underestimated sources of systematic
uncertainties. Thus, the uncertainties of the 1st-grade data were taken from
the original publications,
\item
the 2nd (medium) grade was assigned to a dataset if there were
some omitted or underestimated sources of systematic uncertainties
that could be deduced. Thus, enhanced uncertainties were assigned
to the 2nd-grade data,
\item
the 3rd (lowest) reliability grade was assigned to a dataset if the analysis
revealed defects leading to large and unpredictable systematic errors.
The 3rd-grade datasets were removed from further consideration. Data
remeasured or reprocessed by their author(s) in later publications were also
not considered.
\end{itemize}

The reliability grade of a dataset was lowered if an original publication does
not contain a detailed description of the experimental technique.

The highest grade was assigned, as a rule, to the datasets obtained by modern
techniques: ionization chambers (for the (n,f) reactions),
semiconductor detectors, thin film breakdown counters, solid state
nuclear track detectors with good geometrical conditions. On the other
hand, the datasets obtained by the nuclear photo-emulsion technique
or by fission product yield summation generally received the lowest grade.

In a few cases, two or more datasets show a similar energy dependence of the
cross section but different absolute values. In these cases, renormalization
based on the most reliable absolute cross sections was performed before
fitting.

A few experiments were done with natural Pb targets.
In order to compare these data with those
for pure $^{208}$Pb targets, a search was performed for experimentas
in which (p,f) cross sections were
measured for Pb isotopes~\cite{shigaev73,okolovich,khodai,flerov}.
Then auxiliary approximations were determined for the
$^{206}$Pb/$^{208}$Pb and $^{207}$Pb/$^{208}$Pb
proton-induced fission cross section ratios (see Fig.~5a, b). The following
expression was suggested:

\newpage

\begin{figure}[h!]
\centerline{
\psfig{figure=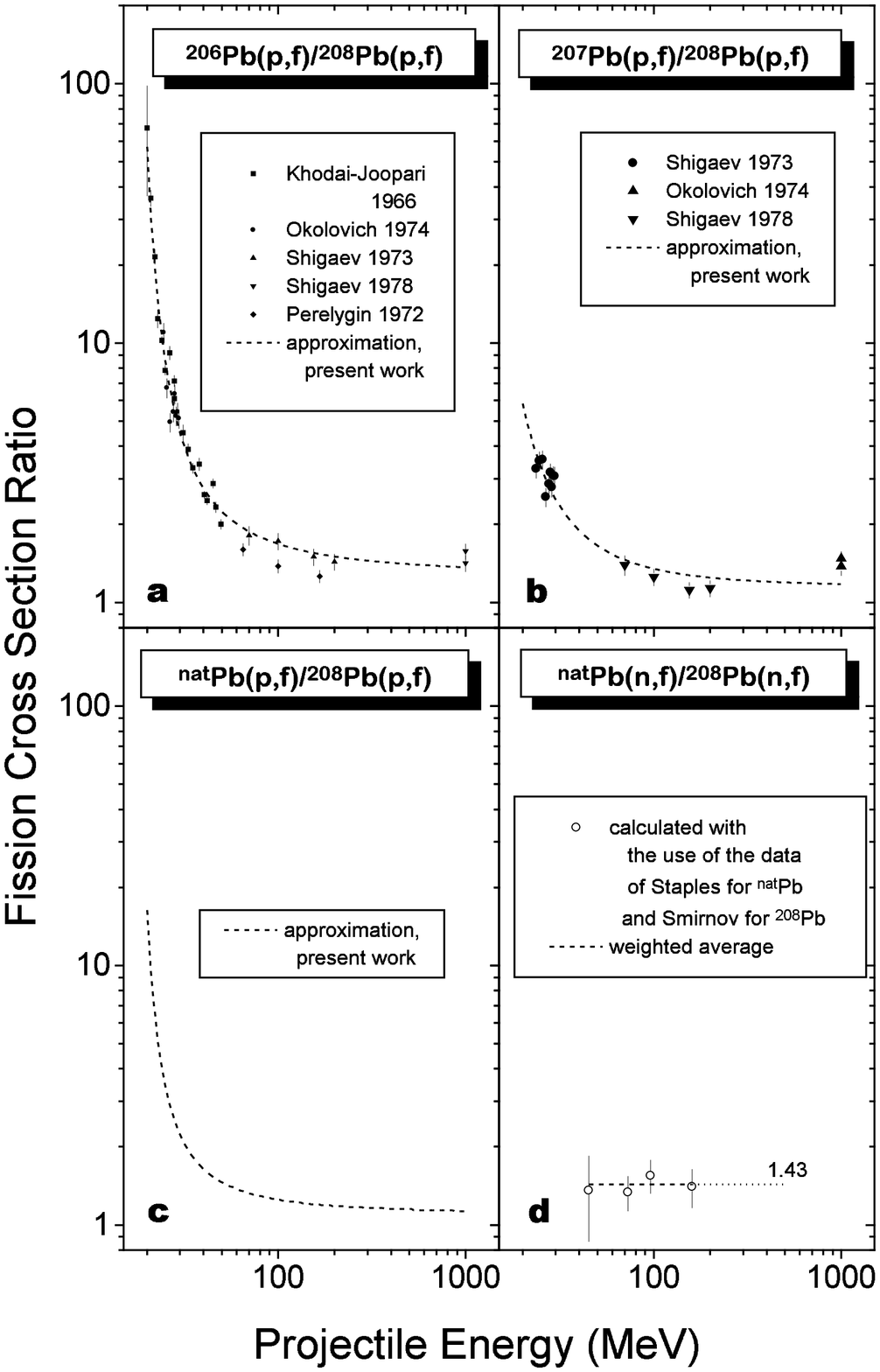,width=145mm,angle=0}}
\end{figure}

{\small
Fig.~5.
The experimental data~\cite{shigaev73,okolovich,khodai,flerov}
and the approximations of the
cross section ratios:
$^{206}$Pb(p,f)/$^{208}$Pb(p,f)~(a),
$^{207}$Pb(p,f)/$^{208}$Pb(p,f)~(b),
$^{nat}$Pb(p,f)/$^{208}$Pb(p,f)~(c),
and
$^{nat}$Pb(n,f)/$^{208}$Pb(n,f)~(d).
}

\newpage

\begin{equation}
r(E)=r_{0}\exp \frac{b}{( \ln E / E_0 )^p} \mbox{ ,}
\end{equation}

\noindent{where $r(E)$
is the cross section ratio, $E$ is the incident particle energy,
$r_{0}$, $b$, $E_{0}$ and $p$ are fitting parameters found by the least-square
method and given in Table III. $\chi^{2} / \nu$ refers to the standard
$\chi^{2}$ per degree of freedom }

\begin{center}
TABLE III\\
The parameters of the $^{206}$Pb/$^{208}$Pb and $^{207}$Pb/$^{208}$Pb
proton-induced fission cross section ratio approximations
\end{center}

\begin{center}
\begin{tabular}{|c|c|c|c|c|c|}
\hline
Cross section ratio & $r_{0}$ & $b$ & $E_0$ & $p$ & $\chi^{2} / \nu$\\
\hline
$^{206}$Pb/$^{208}$Pb & 1.26 & 0.975 & 12.8 & 1.69 & 4.89 \\
$^{207}$Pb/$^{208}$Pb & 1.15 & 16.7  & 3.10 & 3.74 & 3.40 \\
\hline
\end{tabular}
\end{center}

\vspace*{0.5cm}
Combining these approximations with the natural Pb isotopic composition,
we determined an approximation for the $^{nat}$Pb/$^{208}$Pb (p,f) cross
section
ratio (see Fig.~5c). This ratio decreases with incident
proton energy and becomes nearly constant at
an energy of a few hundred MeV.

There are no experiments that include neutron-induced fission cross
section measurements for the relevant separated Pb isotopes within
the same experiment. Thus the $^{nat}$Pb/$^{208}$Pb (n,f) cross section
ratio was calculated with the use of the $^{nat}$Pb
data of Staples et al.~\cite{staples} and
the $^{208}$Pb data of Smirnov et al.~\cite{eismont96,eismont96a}
(see Fig.~5d). The ratio is constant within the
experimental uncertainties and is
equal to $1.43 \pm 0.13$ in the 45--160 MeV energy
region where the data overlap. Assuming that this
ratio weakly depends on energy in the 160--500 MeV region, which is
true for (p,f) reactions discussed above, a renormalization was performed
for the $^{nat}$Pb data of Reut~\cite{reut}, Goldanskiy~\cite{gold55},
and Staples~\cite{staples}.

Figs.~6--9 show the experimental data on the nucleon-induced fission cross
sections for $^{208}$Pb and $^{209}$Bi together with the corresponding
approximations described by the following expressions:

\begin{eqnarray}
\sigma_{f}(E) &=& \exp (c_{0}+c_{1}E+c_{2}E^2) \mbox{ ,} \\
\sigma_{f}(E) &=& \exp [c_{0}+c_{1}\ln E+c_{2}(\ln E)^2] \mbox{ ,} \\
\sigma_{f}(E) &=& c_{0}\left\{1-\exp [-c_{1}(E-c_{2})]\right\} \mbox{ ,}
\end{eqnarray}

\noindent{where
$\sigma_{f}$ is the fission cross section, $E$ is the incident particle
energy, and $c_{0}$, $c_{1}$ and $c_{2}$ are fitting parameters found by the
least-square method and given in Table IV.}

\newpage

\begin{figure}[h!]
\centerline{
\psfig{figure=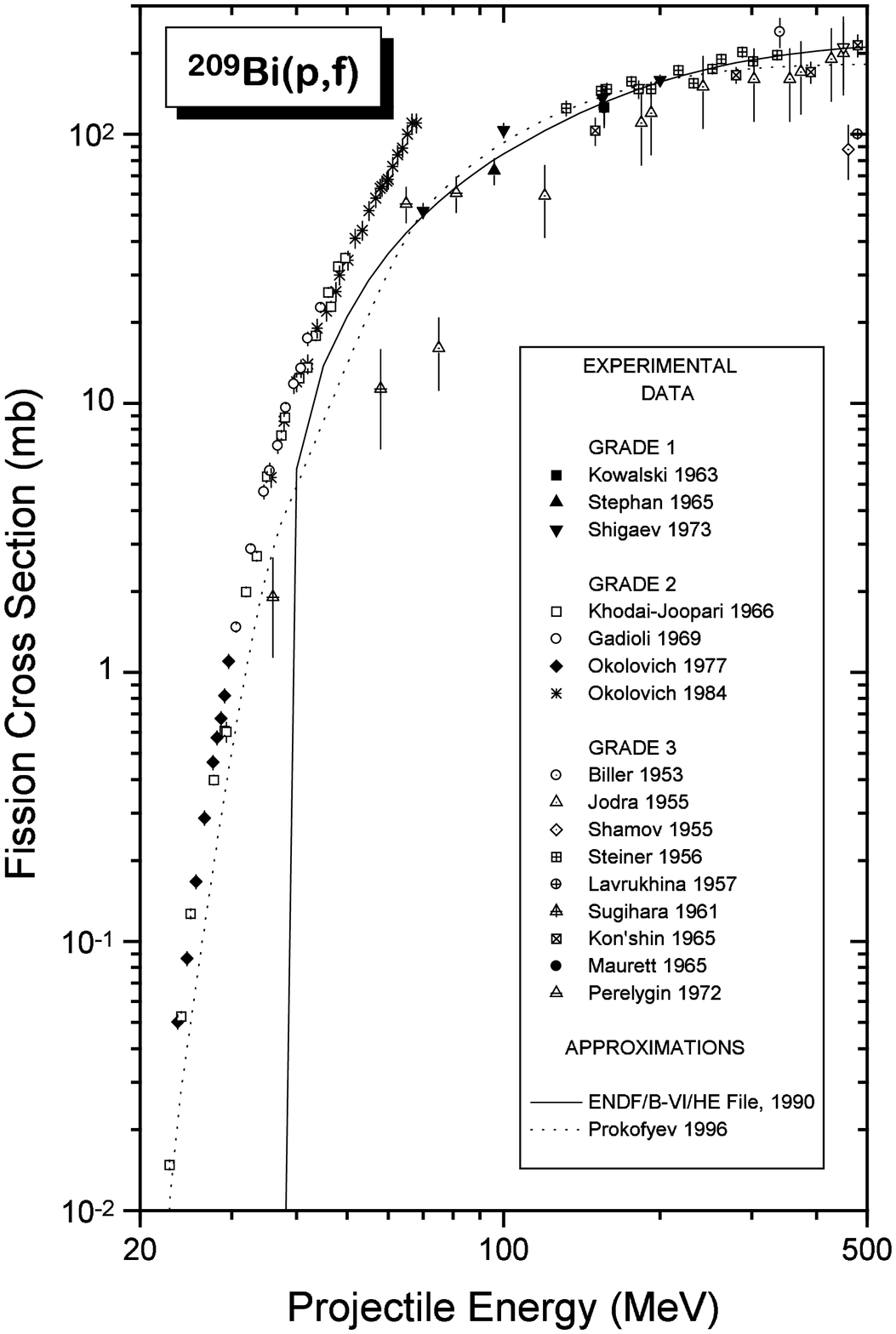,width=145mm,angle=0}}
\end{figure}

{\small
Fig.~6.
The experimental data
\cite{kowalski63}--\cite{shigaev78},
\cite{okolovich},\cite{khodai}--\cite{jodra},
\cite{shamov}--\cite{flerov}
and approximations \cite{sasha97,fukahori91}
for the $^{209}$Bi(p,f) cross section.
}

\newpage

\begin{figure}[h!]
\centerline{
\psfig{figure=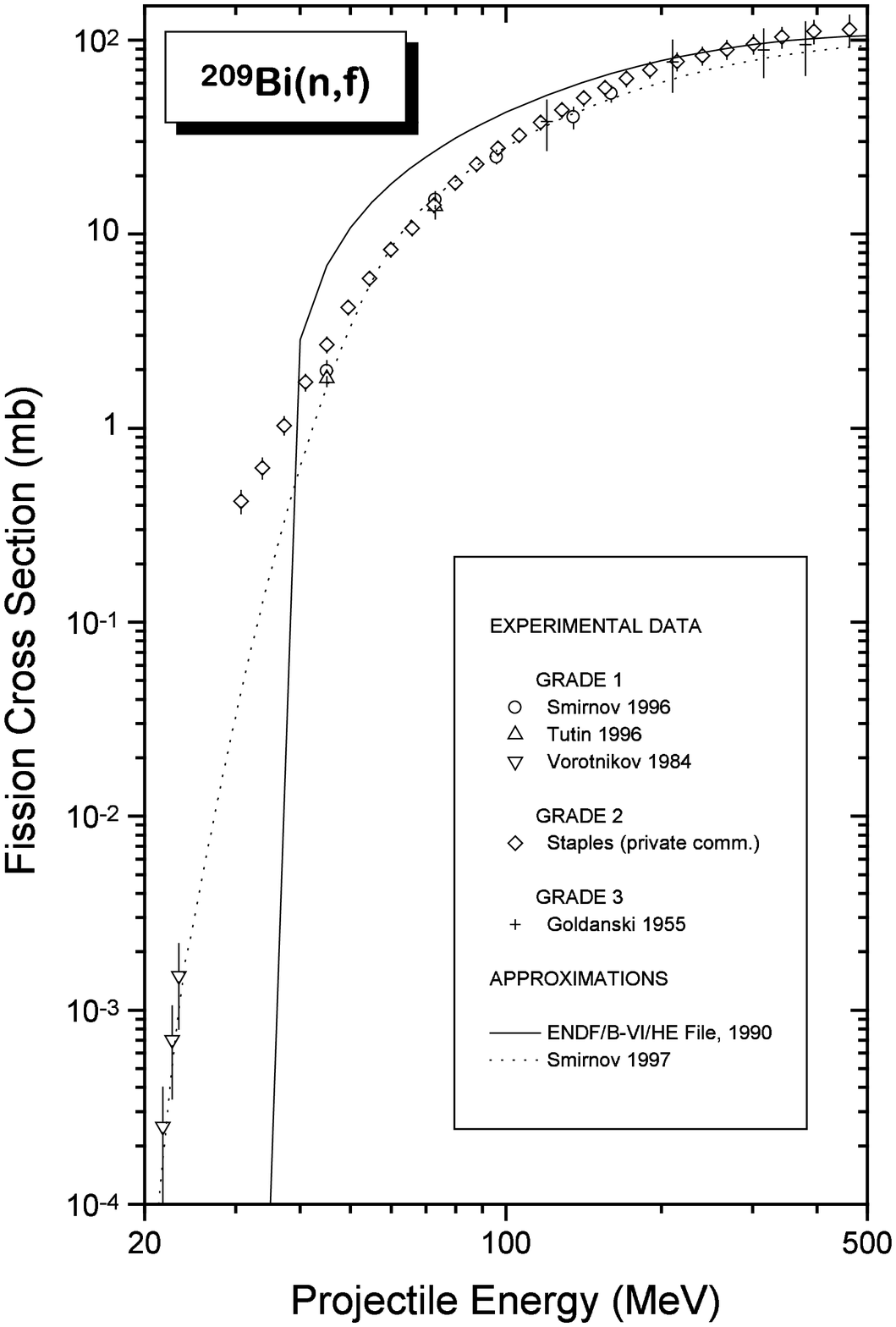,width=145mm,angle=0}}
\end{figure}

{\small
Fig.~7.
The experimental data
\cite{staples,eismont96a,eismont96b,vorotnikov84,gold55}
and approximations \cite{smirnov97,fukahori91}
for the $^{209}$Bi(n,f) cross section.
}

\newpage

\begin{figure}[h!]
\centerline{
\psfig{figure=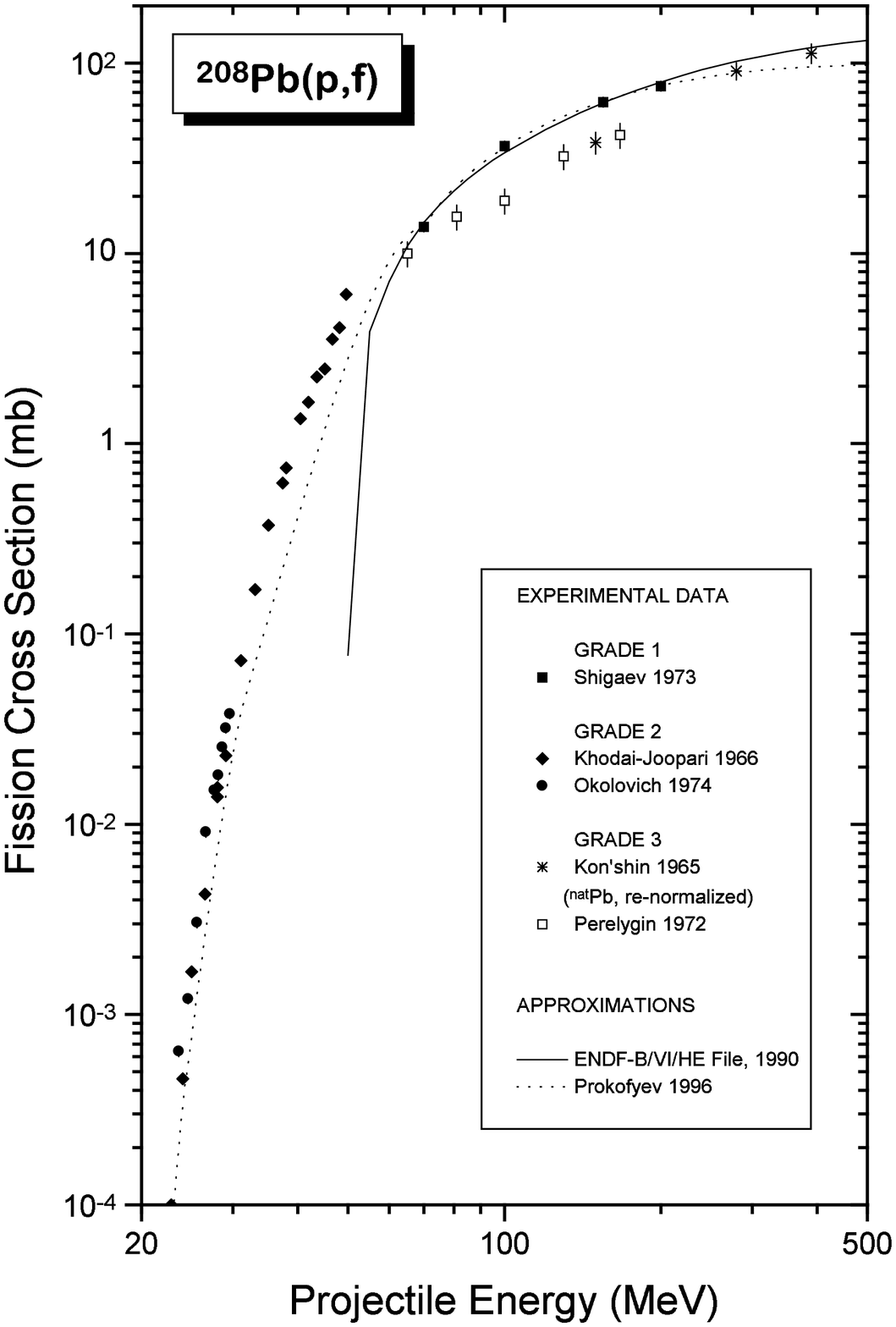,width=145mm,angle=0}}
\end{figure}

{\small
Fig.~8.
The experimental data
\cite{shigaev73,okolovich,khodai,kon66,flerov}
and approximations \cite{sasha97,fukahori91}
for the $^{208}$Pb(p,f) cross section.
}

\newpage

\begin{figure}[h!]
\centerline{
\psfig{figure=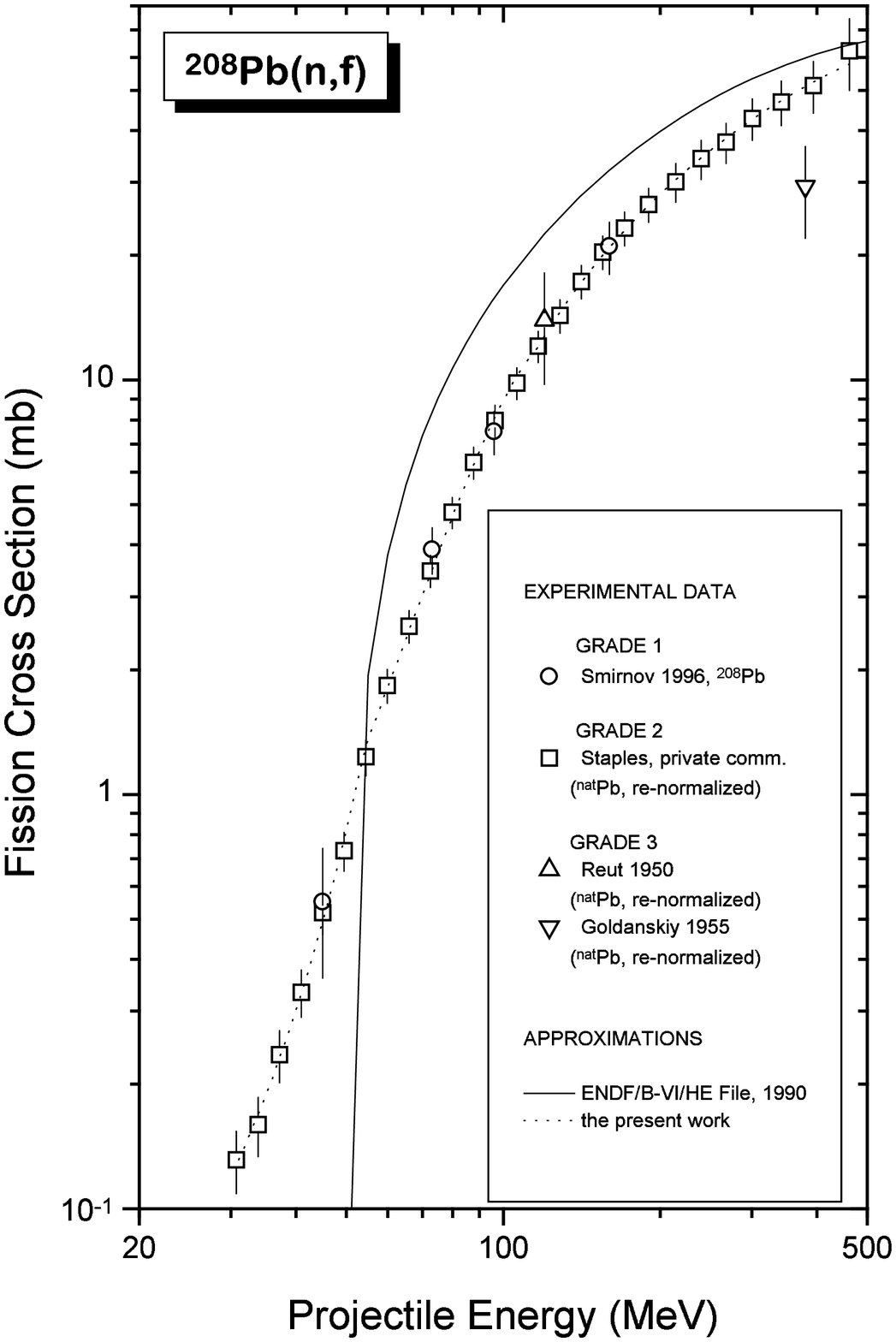,width=145mm,angle=0}}
\end{figure}

{\small
Fig.~9.
The experimental data
\cite{reut,gold55}, \cite{staples}--\cite{eismont96a}
and approximations \cite{fukahori91}
for the $^{208}$Pb(n,f) cross section.
}

\newpage

\begin{center}
TABLE IV\\
The parameters of the nucleon-induced fission cross section approximations for
$^{208}$Pb and $^{209}$Bi
\end{center}

\begin{center}
\begin{tabular}{|c|c|c|c|c|}
\hline
E  (MeV) & Expression & $c_{0}$ & $c_{1}$ & $c_{2}$  \\
\hline
\multicolumn{5}{|c|} {$^{209}$Bi(p,f)} \\
\hline
37.4-70 & (2) & -5.150 & 0.2203 & -0.001290  \\
70-3000 & (4) & 182.7 & 0.01264 & 43.60  \\
\hline
\multicolumn{5}{|c|} {$^{209}$Bi(n,f)} \\
\hline
20-73 & (3) & -108.7 & 50 & -5.6  \\
73-1000 & (4) & 100 & 0.006 & 45  \\
\hline
\multicolumn{5}{|c|} {$^{208}$Pb(p,f)} \\
\hline
31.1-70 & (2) & -16.11 & 0.5310 & -0.003760 \\
70-23000 & (4) & 98.75 & 0.01012 & 55.05 \\
\hline
\multicolumn{5}{|c|} {$^{208}$Pb(n,f)} \\
\hline
30.5--54.5 & (2) & -4.455 & 0.06642 & 3.753*$10^{-4}$ \\
54.5--116.8 & (2) & -4.151 & 0.1026 & -3.920*$10^{-4}$ \\
116.8--462.4 & (4) & 80.71 & 3.190*$10^{-3}$ & 66.06  \\
\hline
\end{tabular}
\end{center}

\vspace*{0.5cm}
The approximations of Fukahori and Pearlstein
\cite{fukahori91} included in the ENDF/B-VI/High
Energy files are also shown in Figs.~6--9.
They are seriously discrepant with respect to the experimental data below
about 40--50 MeV for all four reactions.
In the 50--500 MeV region there are discrepancies of up to a factor of 2
between Fukahori's (n,f) cross section approximations and those given in the
present work, which take into account the recent experimental
data~\cite{staples}--\cite{eismont96b}. The discrepancy for the (p,f) cross
section approximations is less (no more than 35\%).
We consider the approximations given
in the present work as more reliable because they utilize a more complete
experimental database and are obtained as the result of the
consistent procedure of experimental data ranking and selection.\\

\begin{center}
{\large V.
MODELING FISSION CROSS SECTIONS
} \\
\end{center}

The nucleon-induced fission cross section calculations for $^{208}$Pb and
$^{209}$Bi nuclei at 45--500 MeV performed in the present work with the code
CEM95 show that there is no universal input parameter set that allows us to
describe the fission excitation functions in the whole energy region
considered.
In the lower energy part (up to 160 MeV) the best agreement with the
experimental data is reached with the parameter set presented in
Tables V and VI.
Hereinafter the parameter names are given as in the CEM95
code manual~\cite{cem95manual}.

Fig.~10 illustrates the results of the calculations with the given parameters
in comparison with the experimental data approximations discussed in Sect.~IV.

The calculations reproduce the general trends of the
experimental data, although the model systematically underestimates the cross
sections in the middle of the energy range under study (at about 70--100 MeV)
and
overestimates them at the ends (i.e., above 100--150 MeV as well as below
50--60
MeV). An exception is the $^{208}$Pb(n,f) reaction, for which the calculated
results lie under the experimental data for neutron energies below 70 MeV. This
discrepancy may be evidence that the data
approximation for the $^{208}$Pb(n,f) cross section overestimates
the actual cross section at low energies, due to an undetermined systematic
error in Ref.~\cite{staples}.
This approximation was deduced in Sect.~IV by
renormalization of the data of Staples~\cite{staples}
assuming that the $^{nat}$Pb(n,f)/$^{208}$Pb(n,f)
cross section ratio does not depend on neutron energy
(see Fig.~5d). It is possible that this ratio rises with neutron energy
decrease below 50--70 MeV, similarly to the
$^{nat}$Pb(p,f)/ $^{208}$Pb(p,f) cross section ratio (see Fig.~5c),
but the lack of $^{208}$Pb(n,f) cross section data does not allow
us to make a quantitative estimate.

\begin{center}
TABLE V\\
Optimal CEM95 input parameters for nucleon-induced fission cross
section calculations for $^{208}$Pb and $^{209}$Bi targets in
the 45--160 MeV region
\end{center}

\begin{center}
\begin{tabular}{|c|c|}
\hline
Parameter name and & Parameter description \\
recommended value  & \\
\hline
RM=1.5 &
The parameter $r_{0}$ (fm) in Dostrovsky's formula~\cite{dostrovsky} \\
 & for inverse reaction cross sections \\
\hline
WAM  (see Table VI) &
The ratio of the level density parameters in the fission \\
 & and neutron emission channels, $a_{f}/a_{n}$ \\
\hline
IFAM=9 &
The choice of level density parameter systematics $a(Z,N,E^*)$: \\
 & the 3rd Iljinov et al. systematics~\cite{iljinov92} \\
\hline
IB=6 &
The choice of macroscopic fission barrier model $B_{f}(Z,N)$: \\
 & Krappe, Nix, and Sierk~\cite{kns} \\
\hline
ISH=2 &
The choice of shell correction systematics $\delta W_{gs}(Z,N)$ for \\
 & the fission barrier: Truran, Cameron, and Hilf~\cite{cameron70} \\
\hline
IBE=1 &
The choice of fission barrier dependence on excitation energy \\
 &  $B_{f}(E^*)$:Barashenkov, Gereghi, Iljinov, and
Toneev~\cite{barashenkov74}\\
\hline
ISHA=2 &
The choice of shell correction systematics $\delta W_{gs}(Z,N)$ for the\\
 & level density parameter: Truran, Cameron, and Hilf~\cite{cameron70} \\
\hline
IJSP=0 &
The choice of fission barrier dependence on angular \\
 & momentum $B_{f}(L)$: no dependence \\
\hline
\end{tabular}
\end{center}
\vspace*{0.2cm}
\begin{center}
TABLE VI\\
Optimal values for the ratio $a_{f}/a_{n}$ in the CEM95 calculations
of nucleon-induced fission cross
sections for $^{208}$Pb and $^{209}$Bi nuclei in the 45--160 MeV region
\end{center}

\begin{center}
\begin{tabular}{|c|c|c|}
\hline
Target nucleus & \multicolumn{2}{|c|} {Incident particle} \\
  & p & n \\
\hline
$^{209}$Bi & 1.21 & 1.20 \\
$^{208}$Pb & 1.19 & 1.20 \\
\hline
\end{tabular}
\end{center}

\newpage
\begin{figure}[h!]
\centerline{\psfig{figure=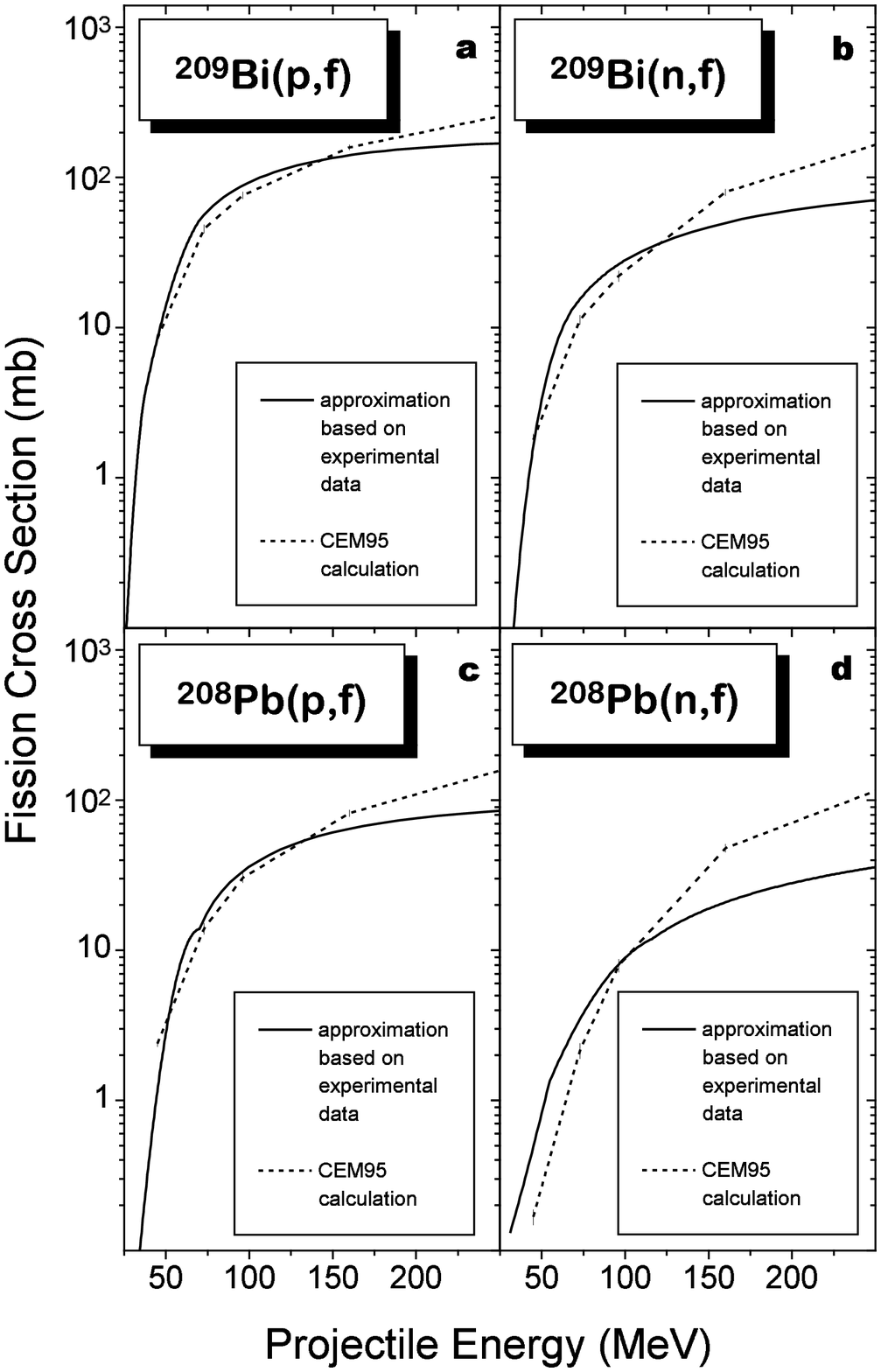,width=140mm,angle=0}}
\end{figure}

{\small
Fig.~10.
Comparison between the
experimental data and the calculations of the cross sections for the
following reactions:
$^{209}$Bi(p,f)~(a),
$^{209}$Bi(n,f)~(b),
$^{208}$Pb(p,f)~(c),
and
$^{208}$Pb(n,f)~(d).
The solid lines represent the approximations of experimental
data given in Sect.~IV, the
dashed lines show the CEM95 predictions with the input parameters shown in
Tables V and VI.
}

\newpage

\begin{center}
{\large VI.
SENSITIVITY OF FISSION CROSS SECTIONS
} \\
\end{center}

Figs.~11--15 illustrate a sensitivity study of the calculated $^{209}$Bi(p,f)
cross sections relative to variations of the following CEM95 input parameters:

\begin{itemize}
\item
the ratio of the level density parameters in the fission and neutron emission
channels, $a_{f}/a_{n}$  (Fig.~11),
\item
the choice of the level density parameter systematics $a(Z,N,E^*)$
(Fig.~12a, b),
\item
the choice of the macroscopic fission barrier model $B_{f}(Z,N)$
(Fig.~13a, b),
\item
the choice of the fission barrier dependence on excitation energy
$B_{f}(E^*)$ (Fig.~14a),
\item
the choice of the fission barrier dependence on angular momentum
$B_{f}(L)$ (Fig.~14b),
\item
the choice of the shell correction systematics $\delta W_{gs}(Z,N)$ for the
fission barrier (Fig.~15a),
\item
the choice of the shell correction systematics $\delta W_{gs}(Z,N)$
for the level density parameter (Fig.~15b).
\end{itemize}

\noindent{In all cases, all other parameters are held fixed as each
is varied in turn. The calculational uncertainties shown in the
figures are due to the statistical character of the Monte Carlo
simulation.}

As seen from Fig.~11, the calculated cross sections are very
sensitive to even
small variations of the $a_{f}/a_{n}$  parameter. The variation of the
$a_{f}/a_{n}$ value by $\pm$ 0.03 from the
optimal value of 1.21 (see Table VI), about 2.5\%,  makes
the $^{209}$Bi(p,f) cross section vary by a factor of 1.5--2, although the
shape of the excitation function does not change appreciably.

Fig.~12 illustrates the calculations with different level density parameter
systematics \cite{iljinov92}, \cite{malyshev}--\cite{cherepanov80}.
A comparative study of these systematics and further useful
references can be found in Ref.~\cite{acta1}.  The sensitivity to
the choice of the level density parameter systematics
is relatively weak. The majority of these systematics give nearly the same
shape to the fission excitation function (with the other parameters fixed
according to Tables V and VI). The absolute scale of the excitation function
varies
from one model to another. For any model
it is possible to improve the agreement with experiment by adjusting
the $a_{f}/a_{n}$  parameter. However, it is not possible to alter the shape of
the calculated excitation function, which is typically too steep above 100 MeV.
The exception is Malyshev's systematics~\cite{malyshev},
which predicts too
gently sloping an excitation function below 70 MeV. The shape closest to
experiment corresponds to the choice of the 3rd Iljinov et al.
systematics~\cite{iljinov92}.
The same choice was found to be the optimal for the CEM in Ref.~\cite{acta1}.

Fig.~13 illustrates the calculational results with different models of the
macroscopic fission barriers
\cite{kns}, \cite{barashenkov73}--\cite{s86}. A comparison of results
obtained with these different
models for fission barriers to experimental data and further references can
be found in Ref.~\cite{acta2}.
Similarly to the previous case, the excitation function shape is
too steep above 100 MeV and depends weakly on the choice of the fission-barrier
model. The exception is the choice of the constant barrier (curve 8 in
Fig.~13b)
that leads to a
too gently sloping excitation function shape below 70 MeV. The
absolute scale of the excitation function varies typically by a
factor of about  2--3 depending on the choice of the specific
fission barrier model, with the other parameters fixed as given in
Tables V and VI.
Again it is possible to improve the agreement with experiment by
adjusting the $a_{f}/a_{n}$  parameter. The best agreement with experiment
was reached with the model of Krappe, Nix, and Sierk~\cite{kns}.
A similar conclusion was made in Ref.~\cite{acta2}.

A similar situation takes place for the choice of the energy dependence of the
fission barriers (see Fig.~14a). The systematics of
Barashenkov, Gereghi, Iljinov, and Toneev~\cite{barashenkov74}
is found to be the optimal one (see Ref.~\cite{acta2} for details).

The choice of the angular momentum dependence of the fission barriers,
or more exactly, the mode of calculation of the saddle-point moment
of inertia according to \cite{strutinsky}
or \cite{cohen} (see details in Ref.~\cite{acta2})
has also a rather weak influence on our present results (Fig.~14b).
This is due to only a
small angular momentum (of only a few $\hbar$) being
imparted to the nucleus by incident nucleons
in the energy region under study. Slightly better agreement with experiment
is reached with no angular momentum dependence of the fission barriers.

Fig.~15 shows the sensitivity of our results to the choice of the shell
correction systematics
\cite{cameron70,myersswiat,cameron57}
for the fission barriers (Fig.~15a) and for the level
density parameter (Fig.~15b). The shell corrections by
Cameron~\cite{cameron57}  and
Truran, Cameron, and Hilf~\cite{cameron70}
give similar results, although application of the latter
option to the fission barriers (curve 2 in Fig.~15a) gives somewhat better
description of the experiment above 100 MeV. This
can be explained as follows. The
shell corrections from Refs.~\cite{cameron57} and~\cite{cameron70}
differ significantly only for nuclei far from stability. For these nuclei,
as was pointed out in Ref.~\cite{acta2}, the corrections~\cite{cameron70}
give a better agreement between
calculated and experimentally measured fission barriers. Since the fissioning
nuclei become more and more neutron-deficient with increase of incident
particle energy (see Fig.~2), it is natural that Truran, Cameron, and Hilf's
shell
corrections~\cite{cameron70} provide a better description of the fission
cross sections at higher energies.

Results quite similar to the ones presented in Figs.~11--15 were obtained from
sensitivity studies for other reactions as well.
In all cases the model predicts too sharp an increase of the excitation
functions for energies above 100 MeV. The
higher the projectile energy, the larger are the discrepancies between the
experimental data and the calculations (see Fig.~16).\\

\newpage

\begin{figure}[h!]
\centerline{\psfig{figure=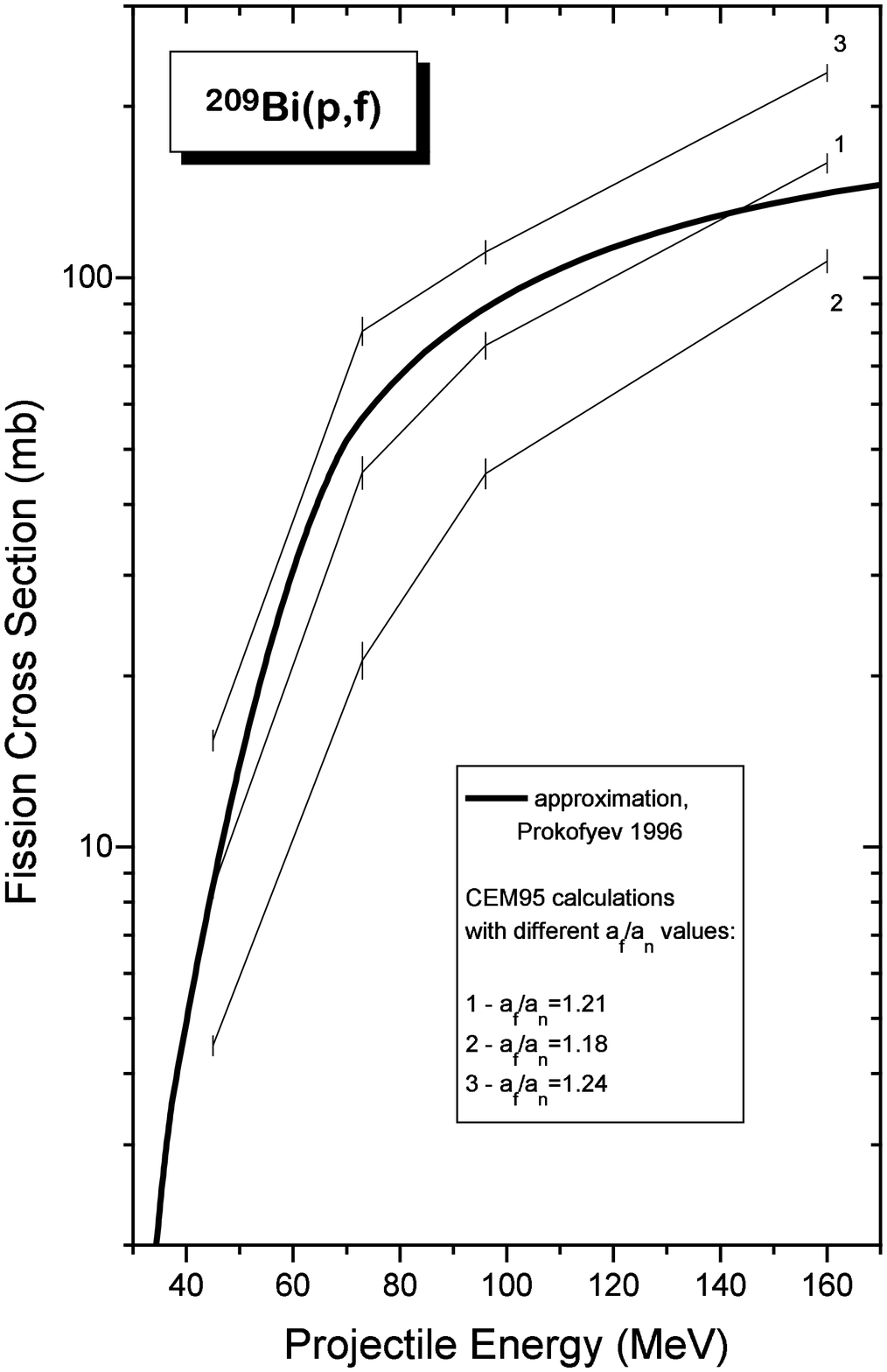,width=140mm,angle=0}}
\end{figure}

{\small
Fig.~11.
The results of the sensitivity study of the
$^{209}$Bi(p,f) cross section calculation with CEM95
to the choice of the ratio of the level density parameters in
the fission and neutron emission channels.
The bold line is the approximation based on the experimental data.
The thin lines with numbers represent the calculations with
different $a_f/a_n$. The other model parameters are fixed according to
Table V.
}

\newpage

\begin{figure}[h!]
\centerline{\psfig{figure=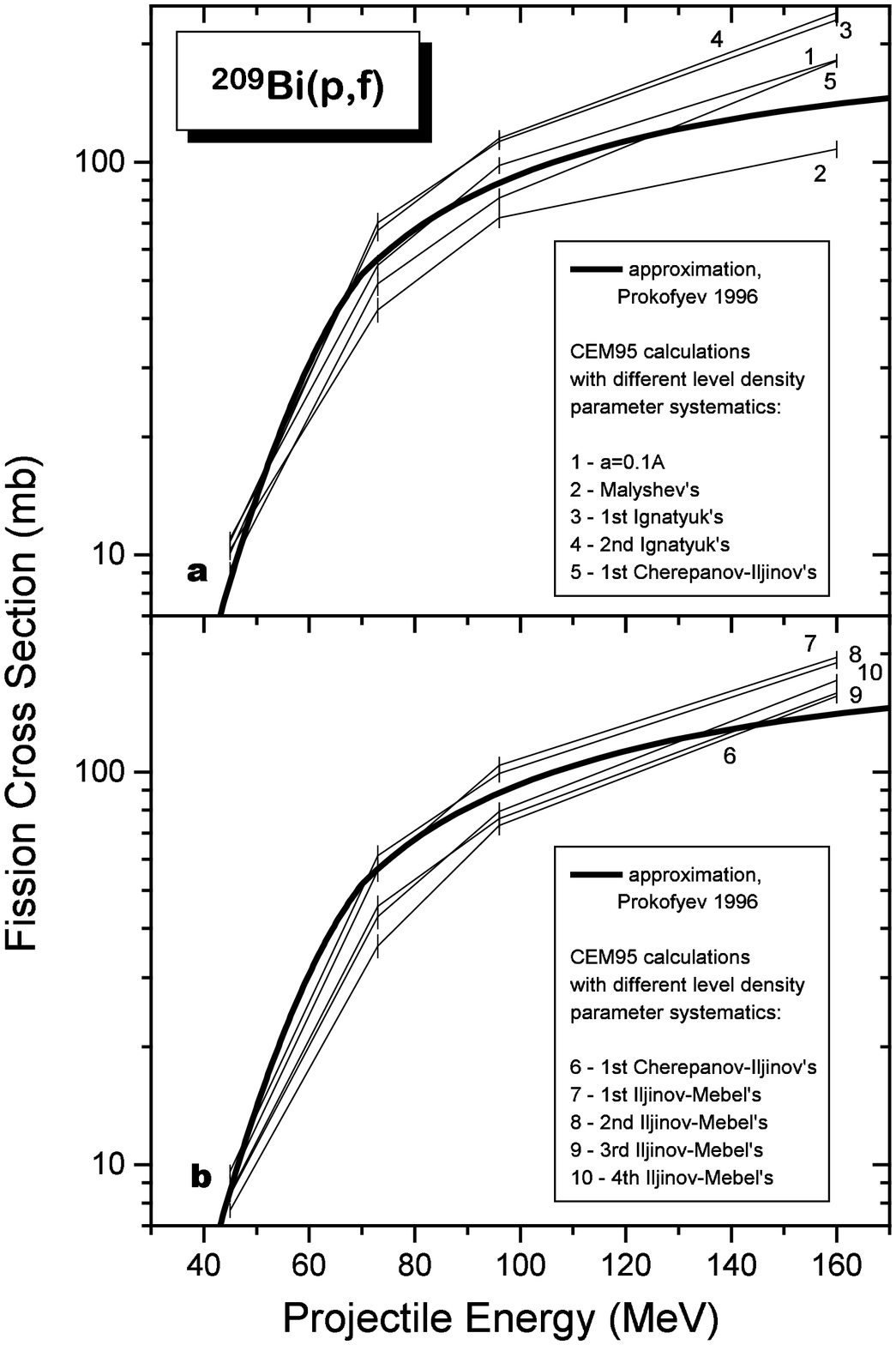,width=140mm,angle=0}}
\end{figure}

{\small
Fig.~12.
The results of the sensitivity study of the
$^{209}$Bi(p,f) cross section calculation with CEM95
to the choice of the level density parameter systematics.
The bold line is the approximation based on the experimental data.
The thin lines with numbers represent the calculations with
different level density parameter systematics
\cite{iljinov92}, \cite{malyshev}--\cite{cherepanov80}.
The other model parameters are fixed according to
Tables V and VI.
}

\newpage

\begin{figure}[h!]
\centerline{\psfig{figure=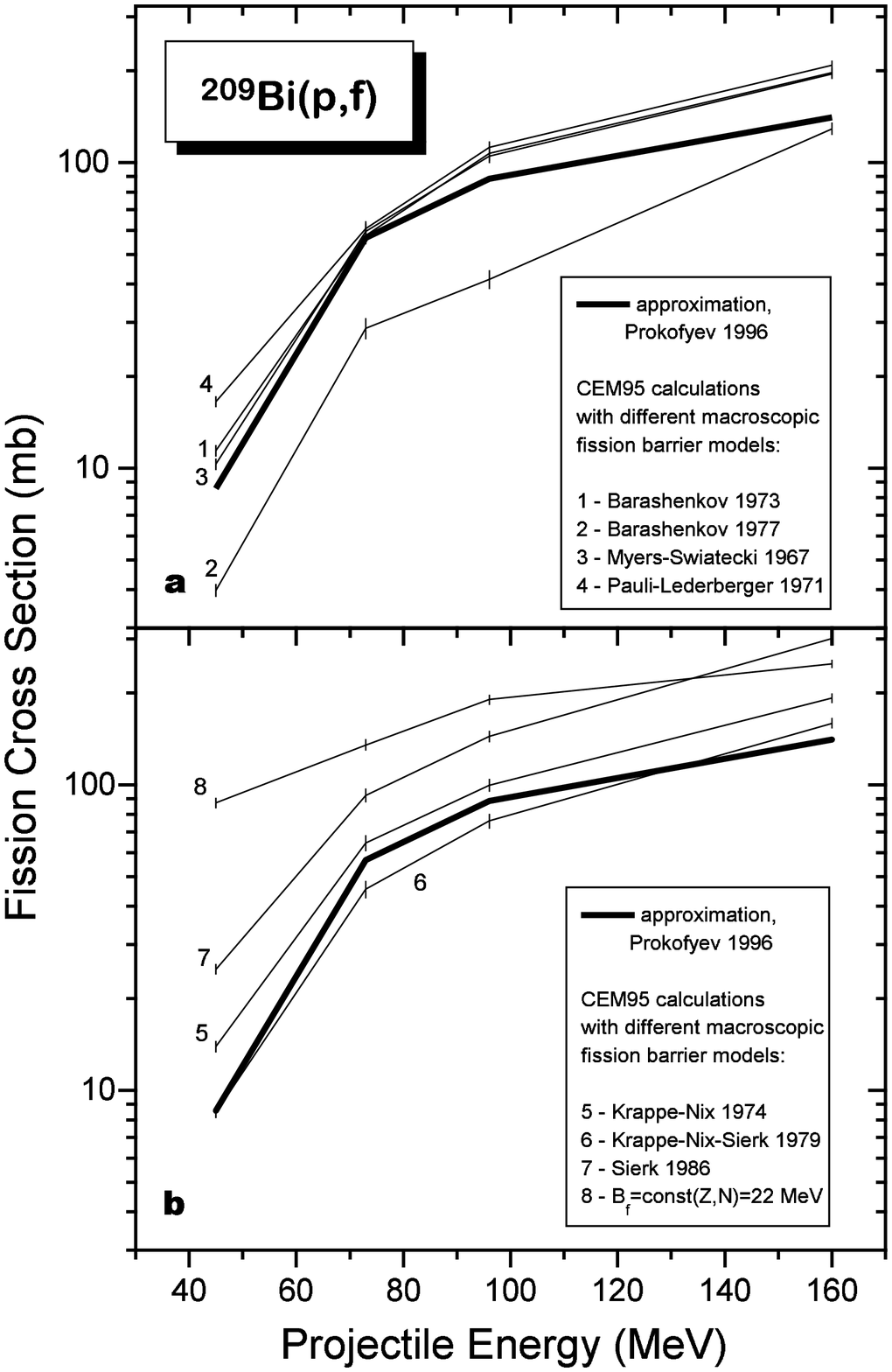,width=140mm,angle=0}}
\end{figure}

{\small
Fig.~13.
The results of the sensitivity study of the
$^{209}$Bi(p,f) cross section calculation with CEM95
to the choice of the model for the macroscopic part of fission barrier.
The bold line is the approximation based on the experimental data.
The thin lines with numbers represent the calculations with
different macroscopic fission barrier models
\cite{kns}, \cite{barashenkov73}--\cite{s86}.
The other model parameters are fixed according to
Tables V and VI.
}

\newpage

\begin{figure}[h!]
\centerline{\psfig{figure=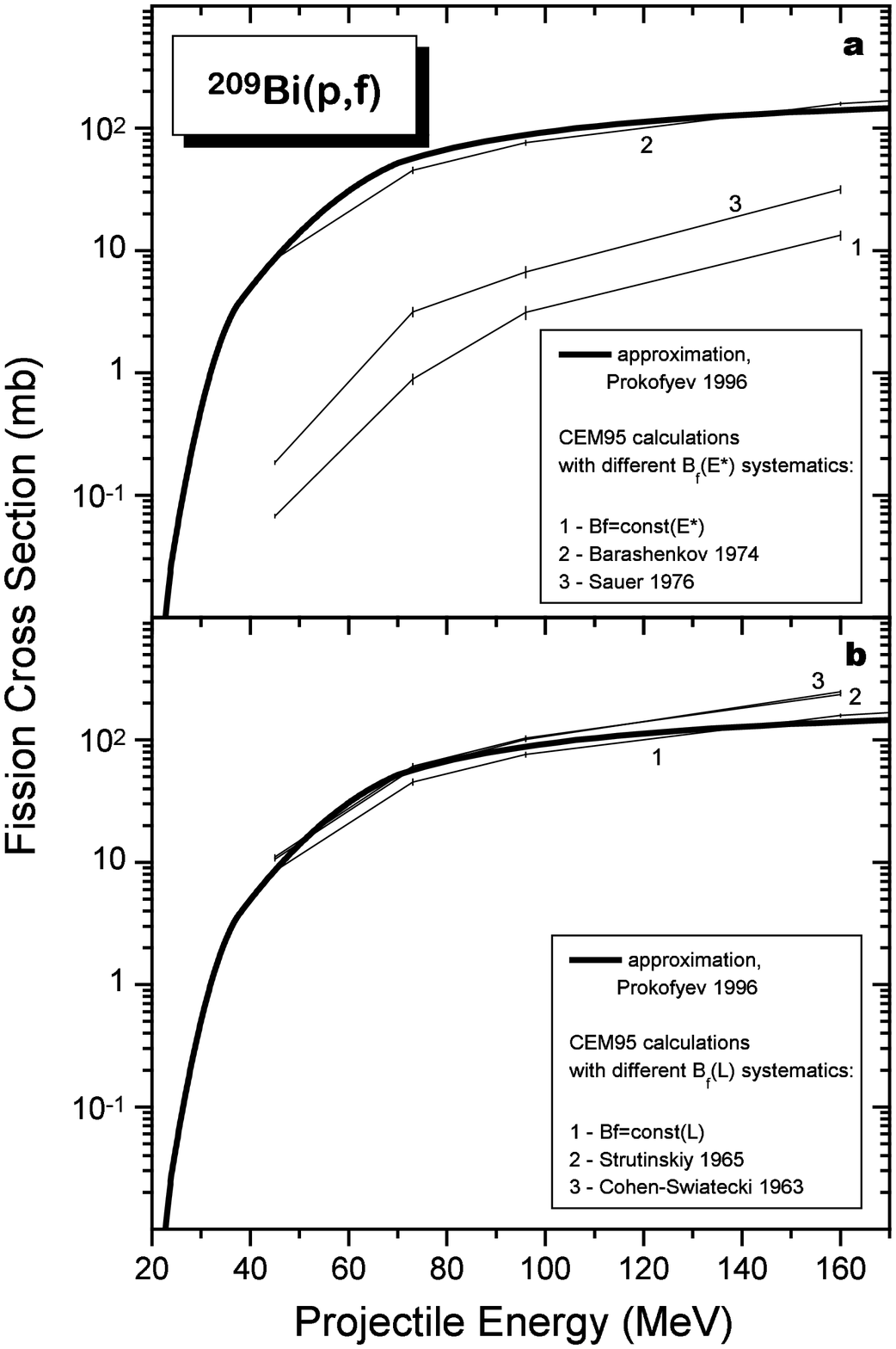,width=135mm,angle=0}}
\end{figure}

{\small
Fig.~14.
The results of the sensitivity study of the
$^{209}$Bi(p,f) cross section calculation with CEM95
to the choice of the systematics for the dependence of fission barrier
on excitation energy~(a) and angular momentum~(b).
The bold line is the approximation based on the experimental data.
The thin lines with numbers represent the calculations with
different $B_f(E^*)$ \cite{barashenkov74,sauer}
and $B_f(L)$ \cite{strutinsky,cohen}
systematics (see \cite{acta2} for details).
The other model parameters are fixed according to Tables V and VI. }

\newpage

\begin{figure}[h!]
\centerline{\psfig{figure=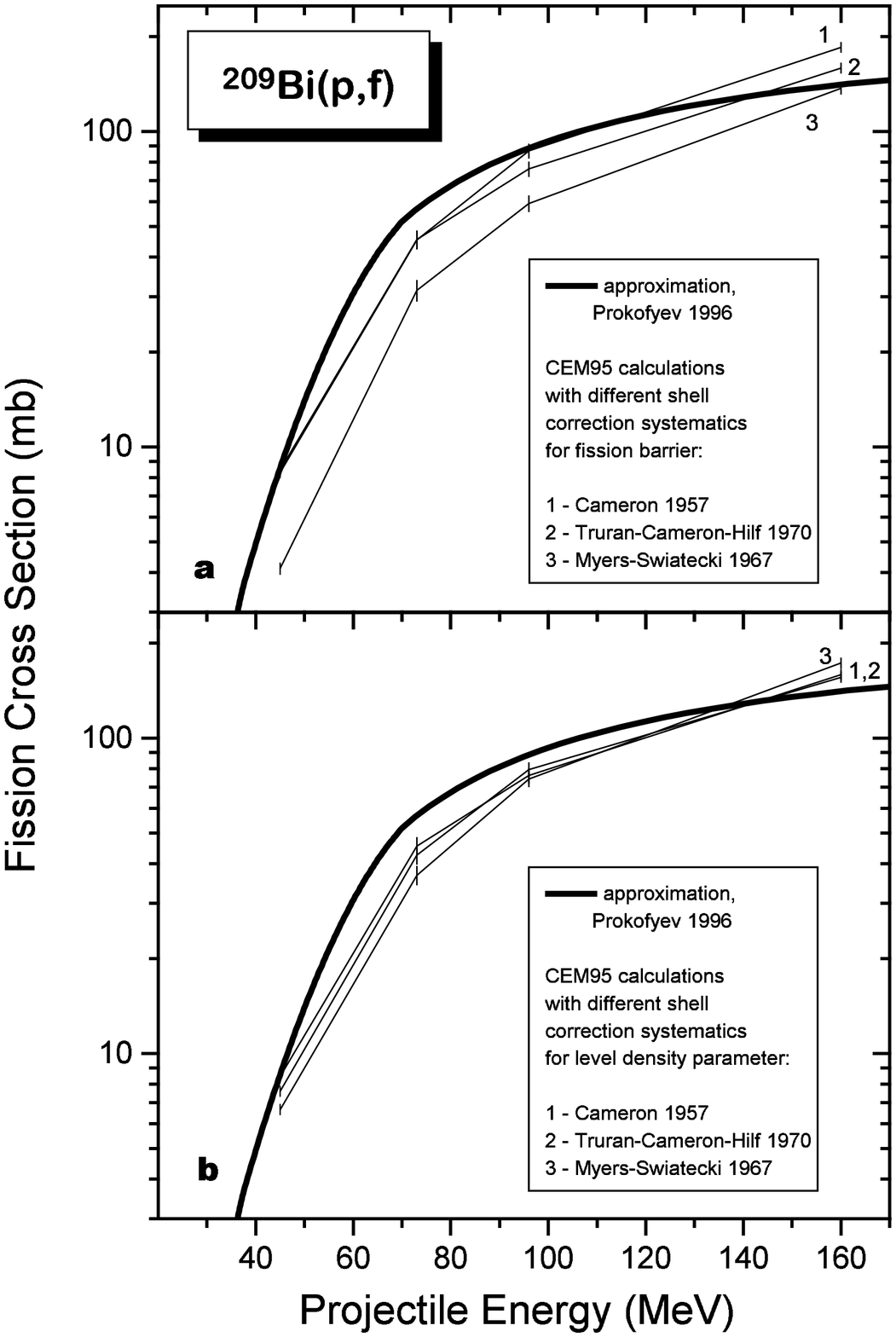,width=135mm,angle=0}}
\end{figure}

{\small
Fig.~15.
The results of the sensitivity study of the
$^{209}$Bi(p,f) cross section calculation with CEM95
to the choice of the shell correction
systematics for fission barrier~(a) and level density parameter~(b).
The bold line is the approximation based on the experimental data.
The thin lines with numbers represent the calculations with
different shell correction systematics
\cite{cameron70,myersswiat,cameron57}
The other model parameters are fixed according to
Tables V and VI.
}

\newpage

\begin{center}
{\large VII. FURTHER DEVELOPMENT OF THE MODEL
} \\
\end{center}

Since CEM95 allows us
to describe well the characteristics of nuclear reactions
that do not involve fission, like double differential spectra of
secondaries (see \cite{cemvar94,nea94a,cemphys}) as well
as spallation product yields (see
\cite{report97}--\cite{mn97}), it is natural to search for the reasons
for the discrepancies in the modeling of fission.
Two different possibilities are:

1) taking into account dynamical effects in fission, which reflect the
connection
between single-particle and collective nuclear degrees of freedom (see, for
example,~\cite{weid89}). The diffusive character of nuclear motion towards
and over the saddle point due to nuclear viscosity leads to a decrease of the
value of the fission width. This effect may rise with the excitation
energy, i.e., in just the direction needed for a better description of the
experimental fission cross sections;

2) modification of the calculation of the level density at the saddle
point. This possibility is discussed below.

As discussed previously, CEM95 incorporates several different level density
parameter systematics.
Most of them utilize the following formula for the excitation
energy dependence of the level density parameter, which was suggested first by
Ignatyuk et al.~\cite{ignatyuk2}:

\begin{equation}
a(Z,A,E^*)=\widetilde{a}(A) \left[1+\delta
W(Z,A) \frac{1-\exp[-\gamma (E^* - \Delta )]}{ E^* - \Delta} \right] \mbox{ ,}
\end{equation}

where $\widetilde{a}$ is the asymptotic value of the level density parameter at
high excitation energies
\begin{equation}
\widetilde{a}(A)=\alpha A + \beta A^{2/3} B_{s} \mbox{ .}
\end{equation}

Here, $A$, $Z$, and $E^*$ are the mass, charge, and the excitation
energy of the
nucleus, $\delta W(Z,A)$ is the shell correction,
$\Delta$ is the pairing correction,
$\alpha$, $\beta$, and $\gamma$ are phenomenological constants,
and $B_{s}$  is the
surface area of the nucleus in units of the surface for the sphere of
equal volume.
For nuclei with equilibrium deformation $B_{s} \approx 1$, while
for the saddle point $B_{s} > 1$. Formula (5) reflects the effect of the strong
correlation between the single-particle state density and the shell correction
magnitude for low excitation energies,
and the fade out of the shell effects on the level density for high excitations
\cite{ignatyuk1,ignatyuk2}.

In the original version of CEM95, the level density parameter systematics based
on formulae (5) and (6) are applied for all decay channels of an excited
nucleus except for the fission channel. In the latter case, the level density
parameter at the saddle point $a_{f}$ is calculated using an
analogous parameter for the neutron emission channel, $a_{n}$, and a constant
ratio, $a_{f}/a_{n}$, which serves as a fitting parameter of the model. Thus
the
shell-effect influence on the level density in the
neutron emission channel is automatically conveyed to the level density at the
saddle point. On the other hand, we expect that shell corrections at the
saddle point for nuclei in this mass range should bear no relation to those at
the ground state, due to the large saddle-point deformation, and the
consequent different microscopic level structure near the Fermi surface
{}.
In fact, the shell corrections at the saddle point should
be of much smaller magnitude than those at the ground state for nuclei
in the neighborhood of $^{208}$Pb, due to the greatly reduced symmetry
of the saddle-point shape compared to the spherical ground state
(see also Refs.~\cite{iljinov80,giardina94}).

In order to estimate the importance of taking into
account this effect, we perform calculations with the parameter $a_{f}$
being
{\it energy-independent}, which is equivalent to the disappearance
of the shell-effect influence on the level density
at the saddle point. The magnitude of the parameter $a_{f}$
was calculated using formula (6) with the same values of the coefficients
$\alpha$ and $\beta$ that
were utilized for the other decay channels. The parameter $B_{s}$  was adjusted
to provide the best agreement with the experimental fission cross
section approximations (see Sect.~IV).
All the other model parameters  were the
same as for the calculations described in Sect.~V (see Table V).
The optimal values of the parameter $B_{s}$  are shown in Table VII.

\begin{center}
TABLE VII\\
The optimal values of the parameter $B_s$ for the fission cross section
calculations with the modified version of CEM95
\end{center}

\begin{center}
\begin{tabular}{|c|c|c|}
\hline
Target nucleus & \multicolumn{2}{|c|} {Incident particle} \\
  & p & n \\
\hline
$^{209}$Bi & 1.18 & 1.12 \\
$^{208}$Pb & 1.15 & 1.12 \\
\hline
\end{tabular}
\end{center}

\vspace*{0.5cm}

The values for the parameter $B_s$
are physically reasonable because they are larger than 1, which reflects the
larger deformation of the fissioning nucleus at the saddle point in comparison
with the equilibrium state and they are not far from the value
$B_{s} \approx 2^{1/3}$, which is
expected for the saddle-point configuration of the
preactinides~\cite{ignatyuk85}. However, we do not yet understand why different
values are needed for proton- and neutron-induced fission, and why the values
of $B_{s}$ in the neutron channel are significantly smaller than those
corresponding to the known deformation of saddle-point shapes in this
mass region.

The ratio $a_{f} / \widetilde{a_n}$, where $\widetilde{a_n}$ is the
asymptotic (large $E^*$) level density parameter in the
neutron channel, is fixed by the fitting of B$_{s}$, and is not independently
fitted as was done previously. $\widetilde{a_n}$ is calculated
using formula (6) and the values $\alpha=0.072$ and $\beta=0.257$,
which correspond to the 3rd Iljinov et al. systematics~\cite{iljinov92}.
The ratios  appear to depend only weakly on the nuclear mass. For
the reactions under study their values (given in Fig.~16) are in the range
1.045--1.070, which are consistent with the compilation of
$\widetilde{a_f} / \widetilde{a_n}$ values published by Ignatyuk et.
al.~\cite{ignatyuk85}.

The results of calculations with the modified CEM95 are shown in Fig.~16.
We see a much better description of the experimental
fission cross sections in
comparison with the original version for nucleon energies of 100--500 MeV. On
the
other hand, the calculation systematically overestimates
to a slight extent the fission cross
sections below 100 MeV. We emphasize that other related improvements
need to be made before a final assessment of the model's value and
predictive capability can be made.  For example, an expected excitation-
energy dependence of the ground-state shell correction would change
the average height of the calculated fission barriers as the incident energy
is increased. However, the current modifications do indicate the crucial
importance
of properly incorporating appropriate level densities and motivate the
search for a consistent model of barriers, ground-state masses, and level
densities which may improve the predictive power of the CEM.

\newpage

\begin{figure}[h!]
\centerline{\psfig{figure=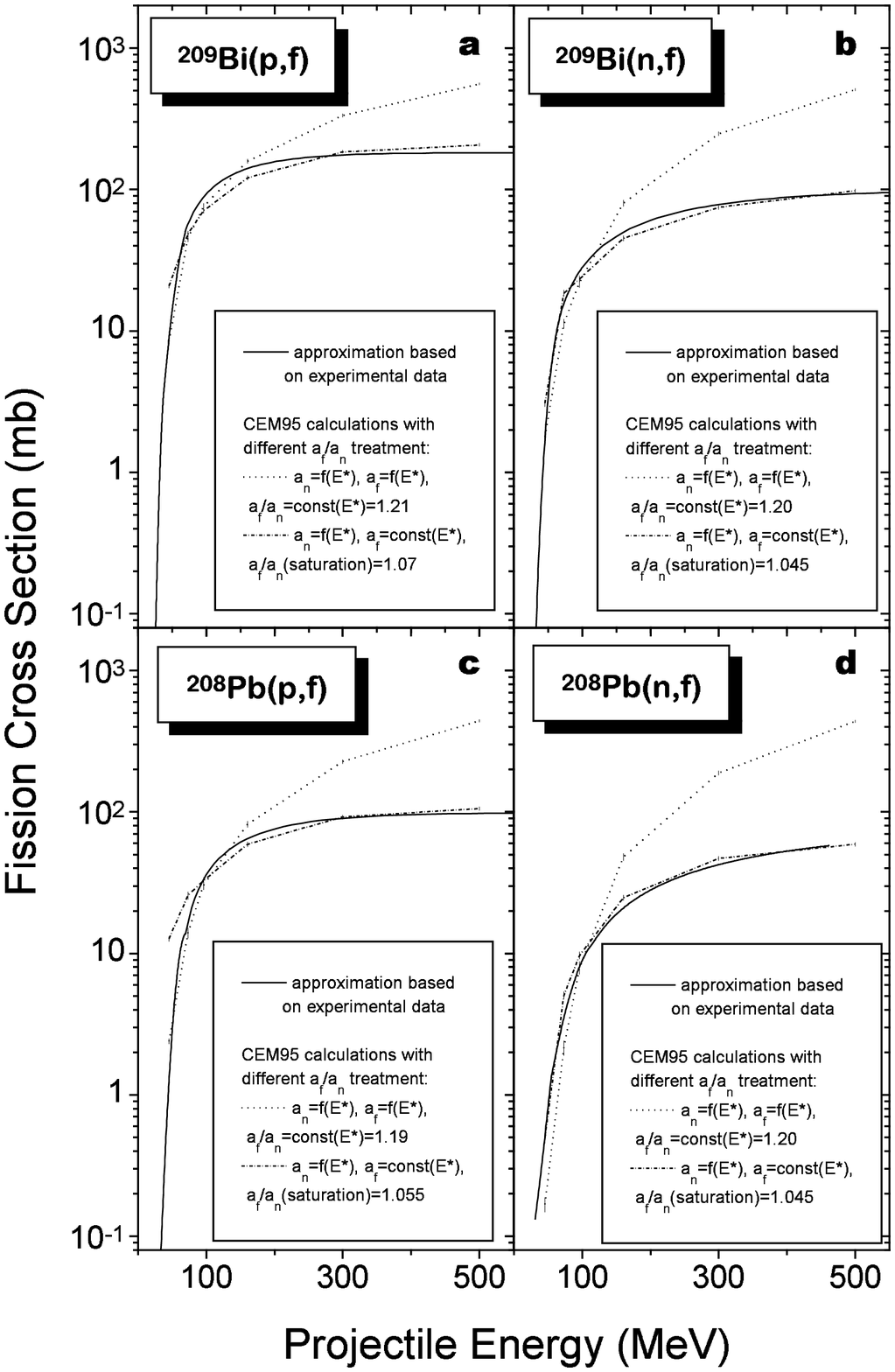,width=140mm,angle=0}}
\end{figure}

{\small
Fig.~16.
Comparison of the experimental data on the
$^{209}$Bi(p,f)~(a), $^{209}$Bi(n,f)~(b), $^{208}$Pb(p,f)~(c),
and $^{208}$Pb(n,f)~(d) cross sections and the calculations with
the original and modified versions of CEM95.
The solid lines represent the approximation of the experimental
data (see Sec. IV).
The dotted lines show the original CEM95 results
with the parameters given in Tables V and VI. The dot-dashed lines show
the results after the modification of the fission channel in CEM95.
}

\begin{center}
{\large VIII. SUMMARY} \\
\end{center}

We have performed a systematic analysis of the nucleon-induced fission cross
sections for $^{209}$Bi and $^{208}$Pb nuclei in the 45--500 MeV energy range
using the CEM95 code,
which has incorporated several choices for the
level density parameters,
nuclear masses, shell and pairing corrections, saddle-point
moments of inertia,
and for the macroscopic and microscopic fission barriers. We have
performed a detailed analysis of the dependence of calculated fission cross
sections and of distributions of the residual nuclei after the cascade
stage of reactions and of fissioning nuclei at the compound-nucleus stage
of reactions on all these choices. We find that
the distributions of residual and fissioning nuclei with respect to their
mass, charge, and excitation energy are not very sensitive to the specific
model
choices used in our calculations. Also,  our results
agree satisfactorily with the scanty data available in the literature
for the average share of the linear momentum transferred to the fissioning
nuclei.

We present here analytical approximations determined by a critical analysis
of all available experimental data,
and compare them to calculations using CEM95 with and without modifications.
Our analysis shows that by choosing the appropriate value for the
ratio of the level density parameters in the fission and neutron
emission channels $a_f/a_n$, it is possible to describe satisfactorily
fission cross sections of all reactions under investigation in a limited
range of energy. But it is impossible to get a satisfactory
agreement in the whole range of energy using a fixed set of CEM95
parameters and of nuclear models used in the calculations. This means
that the version of the CEM as realized in the code CEM95 does not
allow us to predict arbitrary fission cross sections in a large range of
incident energies. We perform in the present
work a further development of the model, where we have modified the
calculation of the level density parameter of nuclei at the saddle point,
by use of simple physical considerations.
This change allows us to get much better agreement with the experimental
fission cross sections in the energy region from 100 to 500 MeV,
but further improvement of the CEM95 is still needed at energies
below about 100 MeV. We plan to continue our work by extending
the range of incident energies and of nuclear targets investigated,
striving for a model capable of predicting fission cross sections for
arbitrary targets in a wide range of incident energies.\\

\begin{center}
{\large ACKNOWLEDGMENTS} \\
\end{center}

It is a pleasure to acknowledge E.~A.~Cherepanov, V.~P.~Eismont,
M.~G.~Itkis, V.~A.~Konshin,
D.~G.~Madland,
P.~M\"oller, J.~R.~Nix, and R.~J.~Peterson
for useful discussions on fission physics.

This study was partially supported by the U.~S.~Department of Energy.

\end{document}